**Water Production Rates from SOHO/SWAN Observations of Six comets: 2017-2020**

Short Title: SOHO/SWAN Observations of comets 2017-2020

M.R. Combi[1], Y. Shou[1], T. Mäkinen[2], J.-L. Bertaux[3], E. Quémerais[3],
S. Ferron[4], and R. Coronel[1]

[1]Dept. of Climate and Space Sciences and Engineering
University of Michigan
2455 Hayward Street
Ann Arbor, MI 48109-2143
United States of America
*Corresponding author: mcombi@umich.edu

[2]Finnish Meteorological Institute, Box 503
SF-00101 Helsinki, FINLAND

[3]LATMOS/IPSL
Université de Versailles Saint-Quentin
11, Boulevard d'Alembert, 78280, Guyancourt, FRANCE

[4]ACRI-st, Sophia-Antipolis, FRANCE

**ABSTRACT**

The Solar Wind ANisotropies (SWAN) all-sky hydrogen Lyman-alpha camera on the SOlar and Heliosphere Observer (SOHO) satellite makes daily images of the entire sky to monitor the three-dimensional distribution of solar wind and solar radiation via its imprint on the stream of interstellar hydrogen that flows through the solar system. In the process it also records the distribution of the hydrogen comae of comets. We report here the analyses of six comets originally from the Oort Cloud observed during the 2017-2020 period by SWAN: C/2015 V2 (Johnson), C/2019 Y1 (ATLAS), C/2017 T2 (PanSTARRS), C/2020 F8 (SWAN), C/2019 Y4 (ATLAS), and C/2019 U6 (Lemmon). Of these the nuclei of C/2019 Y4 (ATLAS) and C/2020 F8 (SWAN) both broke up on their inbound orbit before perihelion. The water production rates over the detectable portion of each comet's orbit are determined and discussed in light of the comet's dynamical sub-class of Oort Cloud comets.

**1. INTRODUCTION**

The Solar Wind ANisotropies (SWAN) all-sky hydrogen Lyman-alpha camera on the SOlar and Heliosphere Observer (SOHO) satellite makes daily images of the entire sky to monitor the three-dimensional distribution of solar wind and solar radiation via its imprint on the stream of interstellar hydrogen that flows through the Solar System (Bertaux et al. 1995). SOHO/SWAN continues to operate nominally since launch in December 1995 (Bertaux et al. 1999) and has observed comets even during the initial cruise phase to insertion in its halo orbit around the L1 Sun-Earth Lagrange point. Since then, over 70 comets have been observed, with publication of many sets of water production rates (Combi et al. 2019). The results for comets observed before 2018 have been archived in the NASA Planetary Data System Small Bodies Node (Combi 2017, 2020). Detailed descriptions of the SOHO spacecraft, SWAN instrument and observation methods for comets appear there and in the survey paper of 61 comets by Combi et al. (2019). The method for determining water production rates is described in detail by Mäkinen and Combi (2005). In all of the figures showing water production rates the internal 1-σ uncertainties are larger for the smaller production rate values for two reasons, because the relative interplanetary background is larger compared to the comet flux and because of the contribution of faint field stars that cannot be clearly identified and subtracted. Furthermore, the value of dQ in the tables reflects the positive value of the uncertainty relative to the value of Q but for large uncertainties the negative value is proportional to dQ/Q. Also, large uncertainties do not necessarily reflect the significant of the detections, but the uncertainty can be large due to systematic errors. The general method makes use of the fact that atomic hydrogen is produced by the photodissociation of $H_2O$ and OH. Coma model photodissociation rates, exothermic velocities and partial thermalization via collisions with the ambient heavy molecule coma are given by Combi et al. (2004). Although the current solar minimum is rather deeper than normal, any effect on the hydrogen analysis is extremely small because of the very large 1° pixels of the SWAN instrument. Also, the 30% uncertainty given in the various figure captions due to model parameters and calibration, while mostly systematic, could have a small random component of a few percent from any dependence of molecular lifetimes on variable solar activity or on some random errors in calibration resulting from extrapolating the Lyman-alpha flux from the nearest time of the Sun facing the Earth where the measurements are made to the actual position facing the comet.

In this paper we describe the results and analysis of SWAN observations of six comets observed by SWAN during the period from 2017 to 2020, as a follow-up to our previous paper on nine comets from 2013 to 2016. The comets observed here are: C/2015 V2 (Johnson), C/2019 Y1 (ATLAS), C/2017 T2 (PanSTARRS), C/2020 F8 (SWAN), C/2019 Y4 (ATLAS), and C/2019 U6 (Lemmon). Table 1 gives a summary of the basic orbital attributes of all the comets including the range in heliocentric distance for the observations incorporated into this study.

## 2. Comet C/2015 V2 (Johnson)

Comet C/2015 V2 (Johnson), hereafter V2, was discovered in images taken on 3 November 2015 with the Catalina Sky Survey Telescope (Nazarov et al. 2015; Facchini et al. 2015). Its orbit had an inbound semi-major axis of 40683 au or a reciprocal semi-major axis of $2.458 \times 10^{-5}$ au$^{-1}$, clearly classifying it as a Dynamically New comet using the definition of A'Hearn et al. (1995). This classification is also in agreement with a value of $1.7 \times 10^{-5}$ au$^{-1}$ given in Nakano Note 4323 (http://www.oaa.gr.jp/~oaacs/nk/nk4323.htm). It reached perihelion on 12.33 June 2017 at a distance of 1.638 au from the Sun. It is predicted to be on a hyperbolic orbit outbound, never to return again to the inner planetary region of the solar system, according to the Minor Planet Center.

Comet C/2015 V2 (Johnson) was observed by SWAN on 64 dates from 25 April to 16 July 2017. Despite the coverage of 2-1/2 months, the comet's perihelion distance of 1.637 au means the range of heliocentric distance covered was rather narrow, from 1.758 au before perihelion to 1.702 au after. The observational circumstances and results of calculating the water production rates are given in Table 2. The water production rate plotted as a function of time since perihelion on 12.3 June (UT) is shown in Figure 1. Since the range of heliocentric distance is quite small (1.638 − 1.758 au), a power-law fit of production rate vs. heliocentric distance is less useful for comet V2, however the values of the exponents are −9.8±0.6 before perihelion and −11.3±1.7 after, or among the steepest in the survey of all SWAN comets (Combi et al. 2019). This steepness is also quite surprising for a dynamically new comet.

There has been nothing reported in the refereed literature on this comet. From visible range spectroscopy, Venkataramani and Ganesh (2018) report no molecular emissions (e.g., $C_2$, $C_3$ and CN) or ion emissions ($CO^+$, $N_2^+$) were detected when the comet was at 2.83 au from the Sun. DiSanti (private communication) reports IR observations with ISHELL on IRTF on UT 7 July 2017, five days before perihelion, giving a water production rate of ~1.5 x $10^{28}$ $s^{-1}$, a $CO/H_2O$ ratio of ~2%, and a rotational temperature of 33±4 K. Jehin (private communication) reported observations of V2 with the TRAPPIST telescope on 2 December 2016 when the comet was 2.26 au from the Sun and the OH production rate was 3.0 x $10^{28}$ $s^{-1}$. They also observed on 28 March 2017 and determined an OH production rate of 2.9 x $10^{28}$ $s^{-1}$ when the comet was 1.92 au from the Sun. The $C_2$ and CN production rates compared with OH were typical.

The result of DiSanti et al. (private communication) for the water production rate from IR spectra is a factor of 2 below the 5.5 x $10^{28}$ $s^{-1}$ found by SWAN on the nearest date, two days later. Also note that the SWAN production rate 13 days earlier, on 27 June, or 11 days before the DiSanti observation, is 6.2 x $10^{28}$ $s^{-1}$, or slightly larger than the later SWAN value. Other than some error in either value, or a coincidental large change in activity down and then back up again during the 13 day period, this could also point to a significant production by an extended icy grain halo around this new comet similar to C/2009 P1 (Garradd) (Combi et al. 2013; Bodewits et al. 2014). In comet C/2009 P1 (Garradd) the ratio of the SWAN production rate and the similar IR observation of Paganini et al. (2012) was a factor of ~3, slightly larger than the ratio for V2. The overall character of the variation of water production over the apparition is also similar to C/2009 P1 (Garradd). See Figure 2. The IR measurement of the rotational temperature of 33±4 K within their aperture of 3-4 arcsec is somewhat higher than would be expected for a coma controlled by normal adiabatic expansion following simple sublimation of water directly from the surface of the nucleus. Their observation was within a 3.5-4 arcsec effective aperture, which at a geocentric distance of 1.04 au corresponds to distances of 2740-3132 km at the comet, which was 1.638 au from the Sun. We have run a coma model (Tenishev et al. 2008; Fougere et al. 2012) accounting only for nucleus sublimation of water and no extended source from sublimating icy grains and calculated the density-weighted rotational temperature of water molecules as a function of aperture size. The result is shown in Figure 3.

A nucleus sublimation-only model would imply a much lower rotational temperature than that observed by DiSanti et al. (private communication) in their aperture. Although coma temperatures typically start near ~200K at the surface and in the Knudsen layer, in the absence of the release of a substantial fraction of water mass in icy grains compared with the mass density of gas from the nucleus, adiabatic expansion quickly decreases the temperature so that by 100 km above the surface the temperature drops to well below 10-20K. This would imply that to maintain a high rotational temperature of 33K at up to 3000 km from the nucleus, a reasonable amount of icy grain heating would be required. For V2 the fraction of the relative grain sublimation source is likely well below that for comet C/2009 P1 (Garradd) or the hyperactive Jupiter Family comets like 103P/Hartley 2 (Fougere et al. 2013) and 73P/Schwassmann-Wachmann 3 (Fougere et al. 2012). The nucleus water sublimation model parameters are as follows:

Coma model parameters:
$Q-H_2O = 4 \times 10^{28}$ s$^{-1}$ (between SWAN and IR results)
$CO/H_2O = 2\%$
$CO2/H_2O = 5\%$ (nominal)
heliocentric distance = 1.678 au

As can be seen in Figure 3, even for an aperture centered on the nucleus of the size used by DiSanti et al., the density-weighted rotational temperature of water is less than 20K. This is likely to be an overestimate because of not accounting for seeing, which lowers the spatial resolution further (Fougere et al. 2012). Therefore, the measured IR rotational temperature, in combination with the production rate being larger for the much larger aperture SWAN measurement than the IR measurement, would be consistent with extra sublimation heating from an extended icy grain halo source.

## 3. Comet C/2019 Y1 (ATLAS)

Comet C/2019 Y1 (ATLAS), hereafter Y1, was discovered in images taken on 16 December 2019 obtained by the ATLAS survey telescope. The comet reached a perihelion distance of 0.837 au from the Sun on 15 March 2020. It had a reciprocal semi-major axis of 0.004693 (or a semi-major axis of 213 au) classifying it as an Old, Long-Period comet (OL) comet using the A'Hearn et al. (1995) dynamical classification scheme, indicating that it has indeed been to the inner solar system before. It has been suggested that comet C/2019 Y1 (ATLAS) is actually the fourth observed fragment of an Old Long Period comet along with C/1998 A1 (Liller), C/1996 Q1 (Tabur) and C/2015 F3 (SWAN). The similarity of the orbital elements suggests they all come from the splitting of a comet with a ~3000-year orbital possibly at a perihelion distance similar to the 0.84 au. (Anonymous reviewer, private communication).

Comet Y1 had a fairly rapid brightening in early January 2020 (Yoshida 2020) as it approached a heliocentric distance of 1 au but then behaved normally for the rest of the apparition. SWAN was able to detect the H coma beginning on March 1, just after the visual magnitude reached $m_v$ ~10 and was detected until early June. The water production rate was determined from the SWAN images throughout this time period (Table 3 and Figure 4). That the water production rate was generally higher for almost 3 months after perihelion than before is also borne out by the asymmetric visual light curve (Yoshida 2020). Otherwise, the water production rate decrease with increasing heliocentric distance after perihelion can be described with a power-law exponent of −2.4±0.1. The very rapid increase before perihelion, when fitted to a power law, gave an exponent of −48±5.

Considering that comet Y1 is not dynamically new, the rapid brightening before perihelion, when it reached a heliocentric distance of ~0.9 au, followed by a typical decrease after perihelion with a power-law slope of −2.4 is noteworthy. Often dynamically new comets have a long, slow, but perhaps higher than normal, increase in activity on the inbound leg of the orbit as essentially fresh surfaces are exposed to more intense solar radiation for the first time (Combi et al. 2019). Comet Y1, on the other hand, had a very low level of activity until the rapid turn-on at a heliocentric distance of about 0.9 au. The asymmetry about perihelion could indicate a highly inclined spin axis and a strong seasonal effect with a change to a new pole about 2 weeks before perihelion. This may be in some ways similar to, but much more extreme than, the behavior of the Rosetta mission comet 67P/Churyumov-Gerasimenko, which has a moderately inclined spin axis and whose north pole is exposed to the sun during the long period during the several years when the comet is far from the Sun but whose south pole is exposed to the Sun during the year or so around perihelion, peaking a month or so after perihelion, and which led to very different evolutions of the nucleus on the two polar regions with much higher levels of both processing and activity by the south polar region (Keller et al. 2015) as well as fallback deposition of material onto the north polar region. Comet Y1, with a semi-major axis of 213 au (and aphelion distance of 425 au), means this comet could likely have had many perihelion passages since leaving the Oort Cloud.

## 4. Comet C/2017 T2 (PanSTARRS)

Comet C/2017 T2 (PanSTARRS), hereafter T2, was discovered on images taken on 2 and 10 October 2017 by the Pan-STARRS survey project (Weryk et al. 2017). It reached perihelion on 4 May 2020 at a distance of 1.615 au from the Sun. It had an original reciprocal semi-major axis of 0.000012, classifying it as Dynamically New (DN), likely on its first trip into the inner solar system. Comet T2 became detectable with SWAN on 1 March 2020 and was detected almost daily until 14 July 2020. Water production rates calculated from the SWAN images are given in Table 4 and are plotted as a function of the time from perihelion in Figure 5. The variation with production rate is slightly asymmetric about perihelion with a somewhat steeper fall-off with heliocentric distance (exponent of $-3.2 \pm 0.1$) than after ($-2.4 \pm 0.2$), though the importance of this is less clear than in some comets because of the narrow range of heliocentric distance covered and the relatively large uncertainties in the determined production rates.

While T2 was visible in the northern hemisphere throughout its apparition, there is little in the published literature about it other than reports of visual magnitudes. See Yoshida (2020). The exception is a brief unrefereed note by Manzini et al. (2020) that reports some concentric structures in broadband images suggesting there may be strong activity driven by rotation with a spin axis along the Earth-comet line. Perhaps this is responsible for the wide erratic spread and dispersion in the day-to-day determinations of water production rates from the SWAN results, which normally smooth out daily fluctuations to some extent.

## 5. Comet C/2020 F8 (SWAN)

Comet C/2020 F8 (SWAN), hereafter F8, was discovered in a SWAN image taken on 25 March 2020 (Mattiazzo 2020). The comet reached a perihelion distance of 0.430 au from the Sun on 27 May 2020. As of this writing the IAU Minor Planet Center has determined it has an eccentricity of 1.0002447, but there is as yet no calculated original reciprocal semi-major axis from the Minor Planet Center. Given its eccentricity, it is likely a Dynamically New comet. Nakano Note NK4152 (http://www.oaa.gr.jp/~oaacs/nk/nk4152.htm) gives a reciprocal semi-major axis of ~ 1 x 10-6 also being consistent with a Dynamically New comet. In addition to the original discovery of F8 in SWAN images, the comet was detected in SWAN images from 1 March 2020 through 7 June. The variation of water production rate determined from the SWAN images is listed in Table 5 along with the observational circumstances. The very unusual production rate variation is plotted as a function of time from perihelion in Figure 6.

There is a break in the production rate results between 9 March and 24 March owing to interference with a very bright star (α Gruis). The production rate (T-78 to T-60 days from perihelion) reached maximum values between 30 April and 3 May of ~$1.2 \times 10^{29}$ s$^{-1}$, which was from 26 to 23 days before perihelion. By perihelion on 27 May the production rate dropped to well below $10^{29}$ and continued to drop precipitously to below $10^{28}$ by 10 days after perihelion, after which it was no longer detectable by SWAN. The visual lightcurve is also unusual and bears out the variation seen in the SWAN results (Yoshida 2020). It is not clear whether the highly irregular and asymmetric activity behavior is due to some seasonal effect or simply that it was a small dynamically new comet that simply reached maximum activity level at ~0.7 au before perihelion (0.43 au), broke up, and then simply faded, like several other dynamically new comets, for example, C/2017 S3 (PanSTARRS) (Mäkinen et al. 2001), 1999 S4 (LINEAR) (Combi et al. (2019b), or like comet C/2019 Y4 (ATLAS), which will be discussed in the next section.

### 6. Comet C/2019 Y4 (ATLAS)

Comet C/2019 Y4 (ATLAS), hereafter Y4, was discovered in images taken on 28 December 2019 made by the ATLAS survey telescope. It reached a perihelion distance of only 0.253 au on 31 May 2020, but had begun to break up weeks earlier and so it is likely that only a dispersing dust cloud was left for the post-perihelion leg of the orbit. Given the brightness before break up and the very small perihelion distance it was promising to be a spectacular, possibly the Great Comet of 2020. Like F8, the Minor Planet Center has yet to determine an original reciprocal semi-major axis, but given its eccentricity of 1.0013279 it is likely a Dynamically New comet. But, also like F8, it reached peak activity well before perihelion and faded rapidly. For Y4 the breakup and dispersal of the nucleus is documented (Hui and Ye 2020) to have occurred on about 22 March 2020, ~70 days before perihelion. Hui and Ye (2020) also presented evidence based on its orbit that C/2019 Y4 along with C/1844 Y1, the Great Comet of 1844, are both fragments of an original parent comet that possibly entered the solar system some 5000 years ago, given the orbital periods of both comets. The water production rate as a function of time from perihelion is shown in Figure 7 and the numerical results, as well as observational circumstances, are listed in Table 6.

Unlike comet F8, the maximum water production rate of Y4 was only $~3 \times 10^{28}$ $s^{-1}$, and so after breakup of the nucleus it was only detectable by SWAN until 10 days before perihelion. The water production remained somewhat level well after the initiation of breakup. There was initially a drop in production rate at about 60 days before perihelion (1 April), but then it actually started to increase a bit, reaching a maximum of $~3 \times 10^{28}$ $s^{-1}$ on 4 May, after which it finally was quite uncertain and fell off to below $10^{28}$ by about 10 days before perihelion on 21 May. Because of the weak signal, all of the production rates are highly uncertain, and should all be taken as simply upper limits or order of magnitude estimates after about 6 May when the comet was at ~0.75 au. A water production rate estimated from two SWIFT observations by Venkataramani et al. (2018), taken when the comet was at 1.25 and 1.01 au, was $1.25 \times 10^{28}$ $s^{-1}$, which is comparable to the average of SWAN observations during that time period. The SWAN average values for the two SWIFT periods with average uncertainty are $(1.40 \pm 0.20) \times 10^{28}$ $s^{-1}$ and $(1.10 \pm 0.35) \times 10^{28}$ $s^{-1}$ or $(1.24 \pm 0.31) \times 10^{28}$ $s^{-1}$ combined. This is excellent agreement with the SWIFT values of $1.25 \times 10^{28}$ $s^{-1}$.

It is apparent that while the nucleus broke into many pieces beginning around 22 March, the pieces were still rich enough with volatile ices to allow sublimation of water for an extended period of time as the fragments continued to get closer to the Sun but dispersed spatially so there was no longer a strong central concentration (Ye et al. 2020a, 2020b). This is different from perhaps the more famous comet to disintegrate, C/1999 S4 (LINEAR), that broke up around the time of its perihelion of ~0.8 au, but whose water production rate dropped precipitously by a factor of more than 20 over 2 weeks. The continued sublimation of volatiles from fragments of a comet long after breakup is not without precedent, as production of volatiles were seen in fragments from the breakup of comet 73P/Schwassmann-Wachmann 3 in 1995 during its subsequent apparition in 2006 (Gilbert et al. 2015).

# 7. Comet C/2019 U6 (Lemmon)

Comet C/2019 U6 (Lemmon), hereafter U6, was discovered by the Mt. Lemmon near Earth Object survey telescope in an image on 31 October 2019. The comet reached a perihelion distance of 0.914 au on 18 June 2020. With a reciprocal original semi-major axis of 0.00207, it would be classified as an old, long-period comet (OL) in the A'Hearn et al. (1995) dynamical classification, indicating that has been through the inner solar system before. There is nothing reported in the literature about other observations of U6.

SWAN was able to detect comet U6 50 days before perihelion, on 28 April 2020, and then observed it on many days until 42 days after perihelion on 29 July 2020. Water production rates and observational circumstances are given in Table 7, and water production rates are plotted as a function of time from perihelion in days in Figure 8. The production rate rose rather rapidly from 28 April until 20 May when it reached a first maximum of $1.35 \times 10^{29}$ s$^{-1}$. Throughout this time period there were, however, fairly wide swings in activity, down to as low as $5 \times 10^{28}$ s$^{-1}$. After this the production rate decreased through perihelion to $7.43 \times 10^{28}$ s$^{-1}$ and then, except for a large 3-day-long outburst on 3 July, reaching a peak of $1.09 \times 10^{29}$ s$^{-1}$, about a factor of two above the level before and after the outburst, the production rate continued to fall more rapidly until it again reached a level of $10^{28}$ s$^{-1}$ on the last detectable day, 29 July. Because of the U-shaped activity both before and after perihelion, and because the range of heliocentric distances was rather narrow (0.91 to 1.27 au), power-law fits to the production rate both before and after perihelion were basically flat with exponents of $-0.3$ and $+0.1$, respectively, and not very useful to provide any physical insight into the condition or state of the nucleus. There is no evidence in the visual light curve (Yoshida 2020) for the factor-of-two outburst seen in the water production on 3 July.

# 8. Summary

We describe herein the results of the analysis of the observations of the hydrogen Lyman-alpha comae of six long-period comets observed during the 2017-2020 time period by the SWAN all-sky camera on the SOHO spacecraft. As of the submission date of this paper there have been few published observations of most of these comets. They range from old long-period comets to truly dynamically new comets, using the dynamical classification of A'Hearn et al. (1995), keeping in mind that those satisfying the reciprocal semi-major axis criteria for being dynamically new have only a 95% probability of being truly dynamically new and on their first trips to the inner solar system from the Oort Cloud. Here is a summary of the major results for each comet.

C/2015 V2 (Johnson) was observed for almost 3 months in 2017 when SWAN images yielded 64 determinations of the water production rate. The comet had a perihelion distance of 1.638 au, but only a fairly narrow range of heliocentric distance was covered (1.638 - 1.758 au). Over this range a power-law fit to the production rate variation with heliocentric distance yielded an extremely steep variation with exponents of -9.8±0.6 before perihelion and -11.3±1.7 after. Peak water production rates were found on 6 June 2017 and 25 June 2017 of nearly $10^{29}$ $s^{-1}$. At the beginning and end of the time period covered the production rate was ~$10^{28}$ $s^{-1}$. Infrared spectroscopic observations by DiSanti et al. (private communication) indicate a higher rotational temperature than the 20K that would be expected for a nucleus-only water sublimation source combined with adiabatic expansion, as shown with a fully-kinetic DSMC model of the coma. An extended source of water from an icy grain halo source could explain the elevated temperature.

C/2019 Y1 (ATLAS) is an old, long-period comet on an orbit with a semi-major axis of 213 au or possibly along with other previous comets is a fragment of an old long-period comet that passed through the inner solar system at a similar heliocentric distance long ago. Its hydrogen coma was detected by SOHO/SWAN from 1 March 2020 until early June. There was a very rapid increase in brightness beginning when it was first detected and the pre-perihelion power-law exponent of -48 ± 5 is consistent with this. On the post-perihelion leg of its orbit the power-law slope of -2.4±0.1 is more like other comets. Since this comet has been through the planetary region of the solar system before, the extreme rapid onset and increase of activity before perihelion is not likely to be due to first exposure of a new surface but rather a seasonal effect, similar to, but much more extreme than, that of the Rosetta mission comet 67P/Churyumov-Gerasimenko, whose south pole faces the Sun only on its several months around perihelion but the north pole faces the Sun on the long distant part of its orbit.

C/2017 T2 (PanSTARRS) is classified as a Dynamically New comet, likely on its first trip into the inner solar system. SOHO/SWAN detected the hydrogen coma from 1 March through 14 July 2020, reaching a perihelion distance of 1.615 au on 4 May. The water production rate variation over the apparition showed it to be slightly asymmetric about perihelion, having a somewhat faster drop with increasing heliocentric distance before perihelion than after. Like some of the other comets in this group there are no published observations of other production rates.

C/2020 F8 (SWAN) was actually discovered on a SWAN image taken on 25 March 2020. It was found in pre-discovery SWAN images starting on 1 March and detected until 7 June. The comet reached a perihelion distance of 0.430 au on 27 May, reaching a maximum water production rate of $1.2 \times 10^{29}$ s$^{-1}$, 26 to 23 days before perihelion. With an eccentricity slightly in excess of 1, it is likely a dynamically new comet, but there is as yet no definitive determination of its original semi-major axis. It had a quite unusual and highly asymmetric activity variation with respect to perihelion over this time with rather irregular variations generally increasing from $2 \times 10^{28}$ s$^{-1}$ to just over $1 \times 10^{29}$ s$^{-1}$ about 25 days before perihelion and then dropping precipitously after perihelion by more than one order of magnitude within 10 days. Although there are no supporting observations, it would appear that the comet disrupted and disintegrated around perihelion, reminiscent of C/1999 S4 (LINEAR), C/2017 S3 (PanSTARRS) and C/2019 Y4 (ATLAS). No visual magnitude determinations were reported by Yoshida (2020) after perihelion.

C/2019 Y4 (ATLAS) was another comet that broke up on this trip into the inner solar system. Its eccentricity would indicate that it is a dynamically new comet but Hui and Ye (2020) have shown that this comet was likely a broken fragment from a comet that passed already — possibly on the real first trip into the inner solar system — once before, but without substantially changing its aphelion. As discussed in A'Hearn et al. (1995) the classification based on the original semi-major axis is only a statistical estimate. Comets with semi-major axes exceeding 20000 au are likely to be dynamically new, perhaps with a 90% probability. The comet's water production rate varied from just over $10^{28}$ s$^{-1}$ before perihelion and varied irregularly between $5 \times 10^{27}$ and $3 \times 10^{28}$ s$^{-1}$. After the nucleus broke up, the production rate decreased slowly, and the last usable SWAN measurement came about 10 days before its 31 May perihelion. The decreasing production rate after breakup indicated that the remaining fragments were icy and continued to produce water for a few weeks after break up after a sharp detectable nucleus disappeared and as what was left of the comet moved closer and closer to the Sun.

C/2019 U6 (Lemmon) is an old, long-period comet, and like some others described here has no species observations reported in the literature. SWAN detected the hydrogen coma beginning on 28 April 2020 lasting until 29 July. The water production rate rose rapidly for 20 days after the first detection and then varied rather widely and irregularly until perihelion on 18 July.

Then, with the exception of a factor of 2 outburst that peaked on 29 July that lasted about four days, the production rate dropped rather normally with increasing heliocentric distance. Because of the rather irregular activity variation and the outburst, power-law fits to the variation with heliocentric distance are not illuminating.

**Acknowledgements:** SOHO is an international mission between ESA and NASA. M. Combi acknowledges support from NASA grant 80NSSC18K1005 from the Solar System Observations Program. T.T. Mäkinen was supported by the Finnish Meteorological Institute (FMI). J.-L. Bertaux and E. Quémerais acknowledge support from CNRS and CNES. We obtained orbital elements from the JPL Horizons web site (http://ssd.jpl.nasa.gov/horizons.cgi). For classification of the dynamical ages of the comets in this paper we used the Minor Planets Center web site https://www.minorplanetcenter.net/db_search tool. The composite solar Lyα data were taken from the LASP web site at the University of Colorado (http://lasp.colorado.edu/lisird/lya/). We acknowledge the personnel who have been keeping SOHO and SWAN operational for over 20 years, in particular Dr. Walter Schmidt at FMI. We also acknowledge the support of R. Coronel by the Undergraduate Research Opportunity Program of the University of Michigan. We also thank the two reviewers for their careful reading and helpful suggestions that have improved the paper.

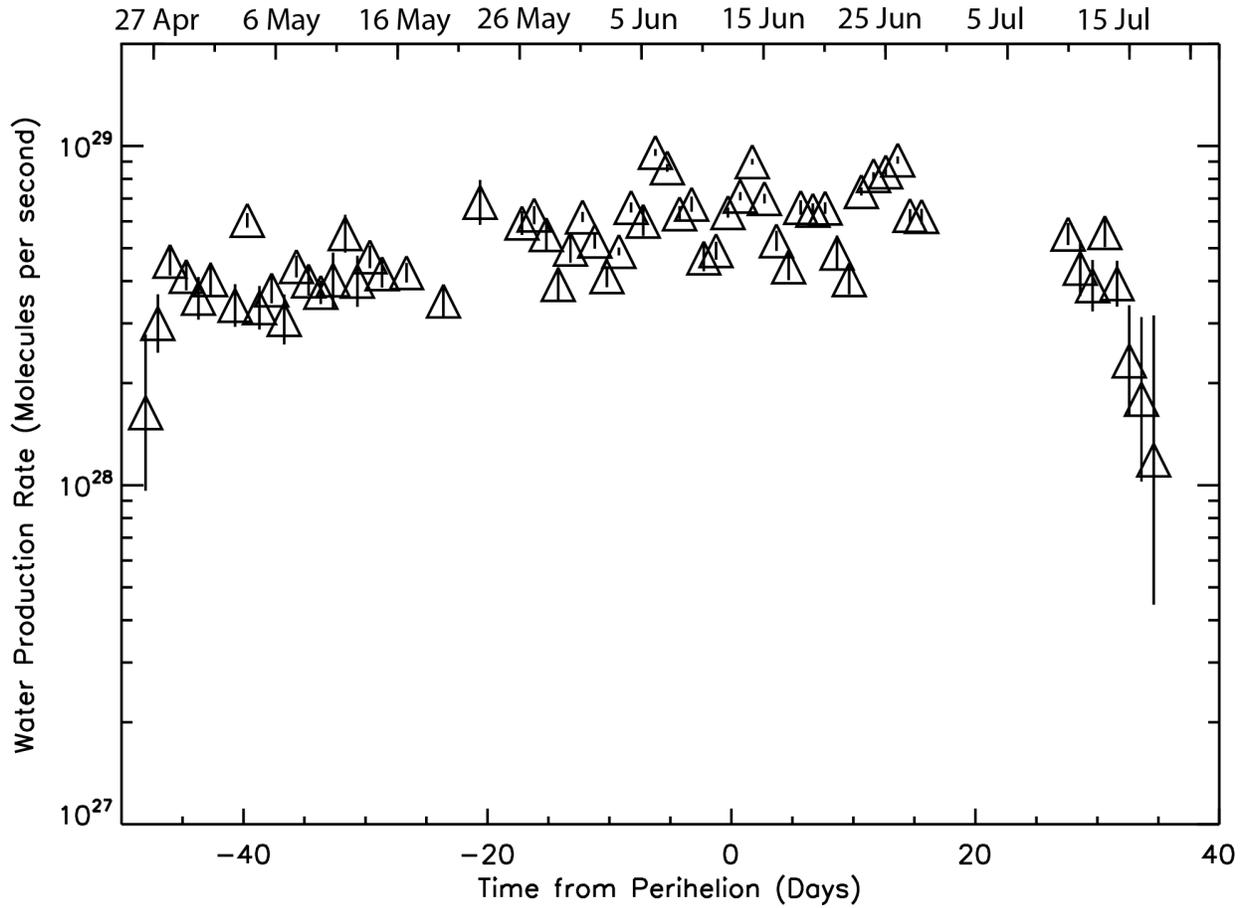

Figure 1. Water production rate as a function of time from perihelion and UT date 2017. The points give the water production rate in $s^{-1}$ from single images. The error bars give the 1-σ formal random fitting errors for each value. There is a ~30% uncertainty from the model parameters and calibration.

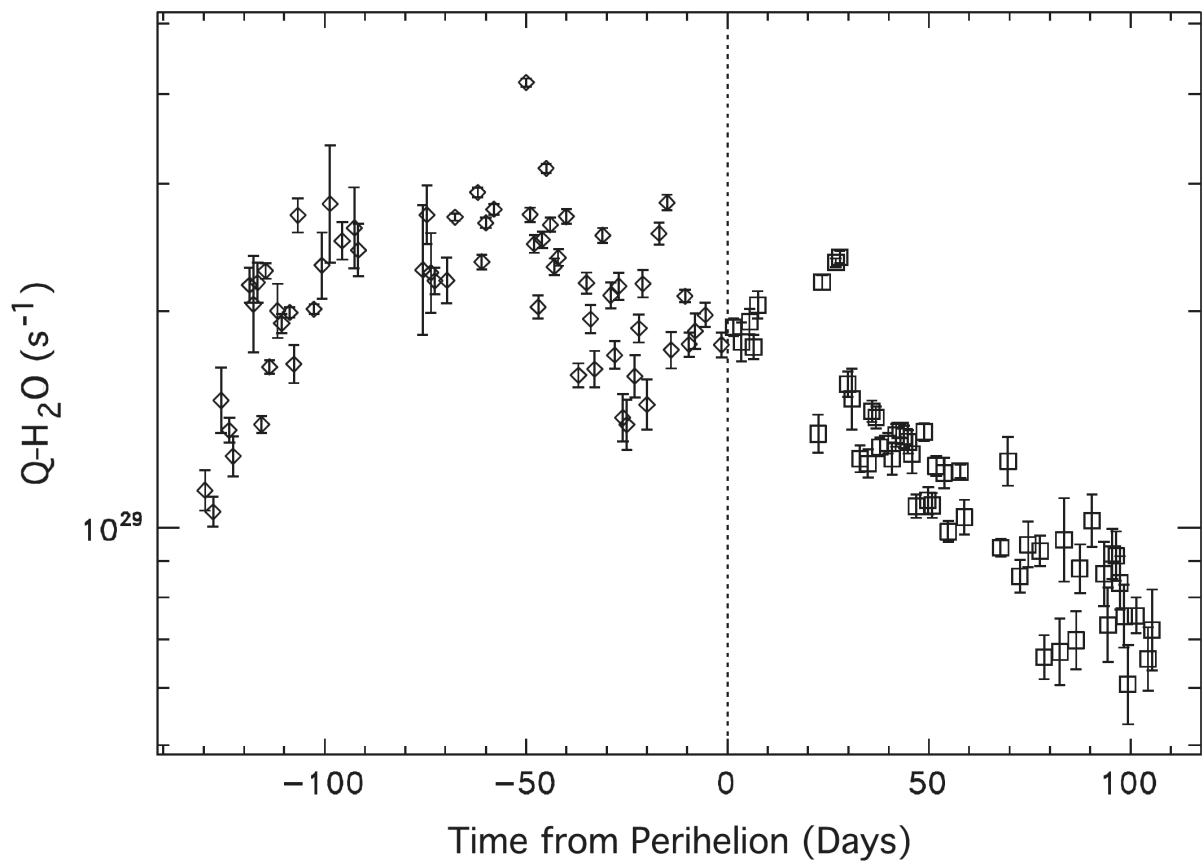

Figure 2. Water production rate in comet C/2009 P1 (Garradd) as a function of time from perihelion. Figure from Combi et al. (2013).

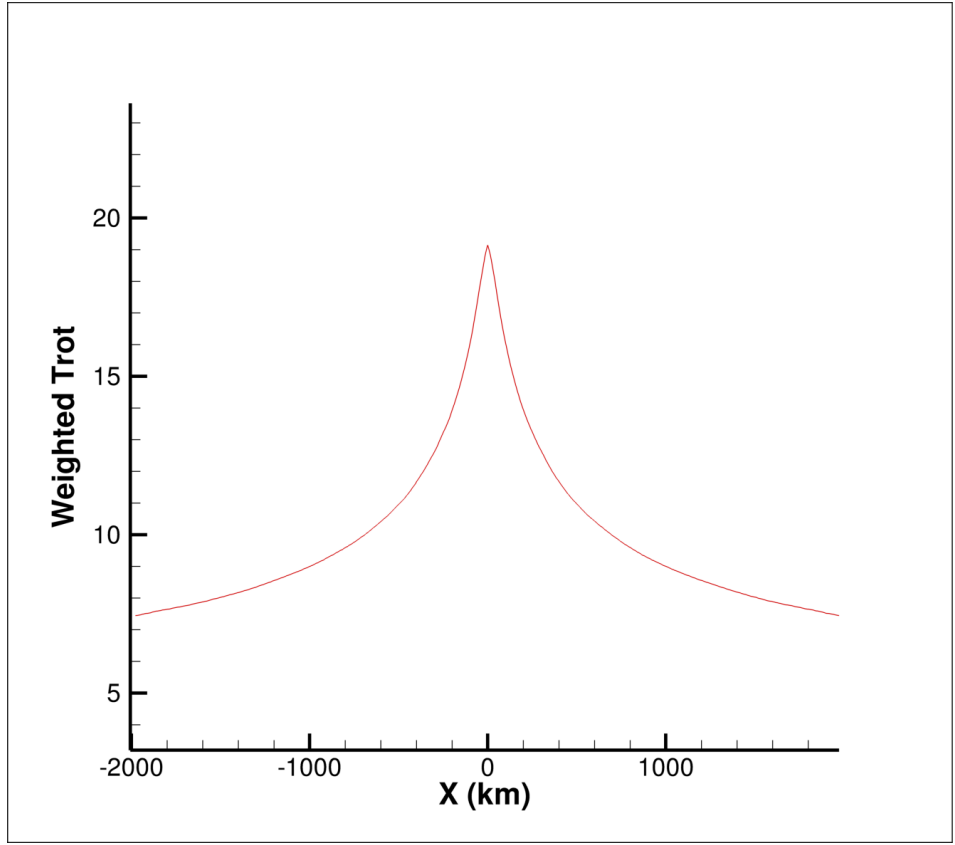

Figure 3. Coma model of the rotational temperature weighted by the water density as a function of aperture size centered on the nucleus. Clearly for an aperture the size of the DiSanti et al. (2020) observation (~3000 km) the weighted Trot is less than 20K, well below the measured value of 33±4 K.

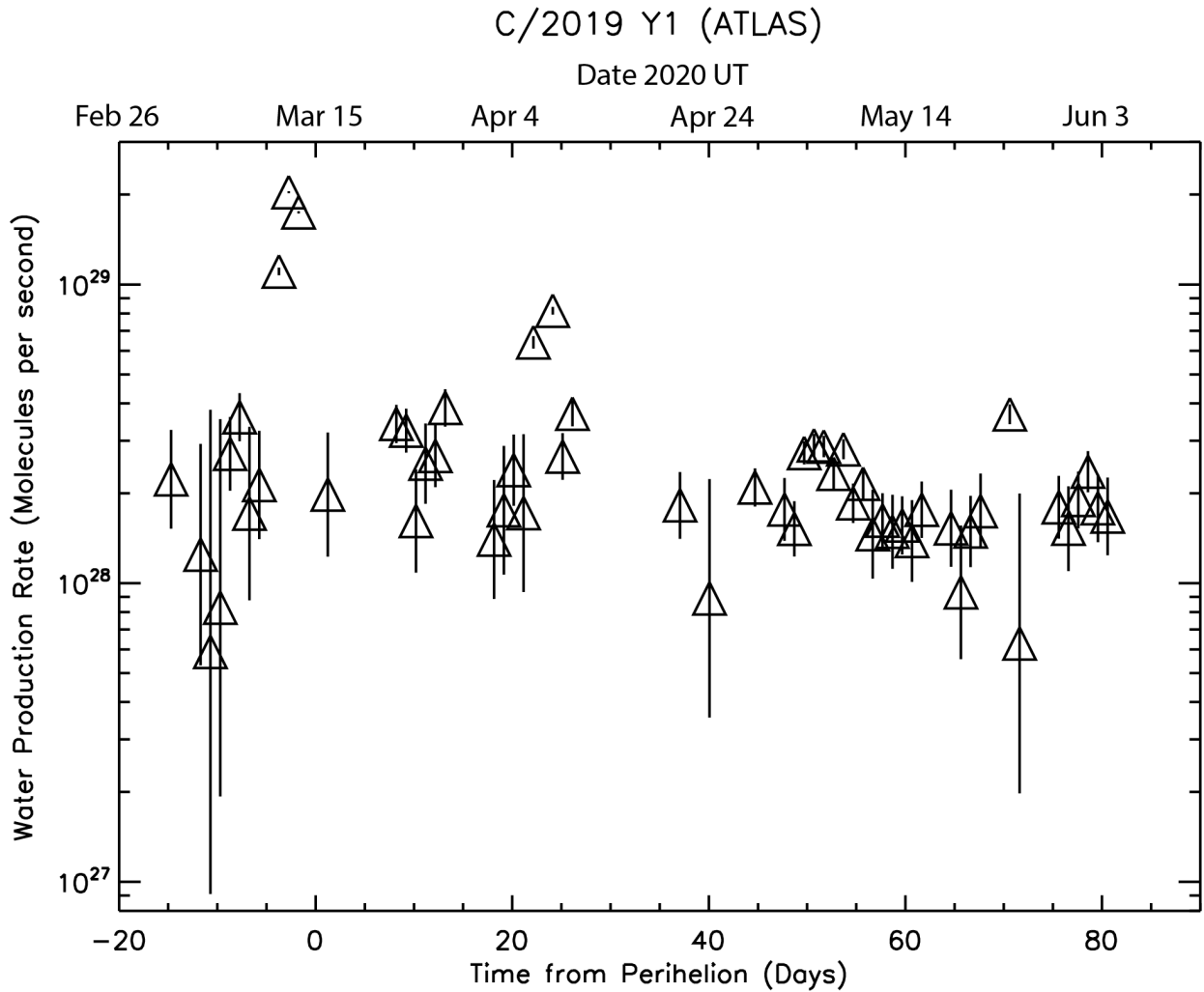

Figure 4. Water production rate in comet C/2019 Y1 (ATLAS) as a function of time from perihelion and UT date 2020. The points give the water production rate in s$^{-1}$ from single images. The error bars give the 1-σ formal random fitting errors for each value. There is a ~30% uncertainty from the model parameters and calibration.

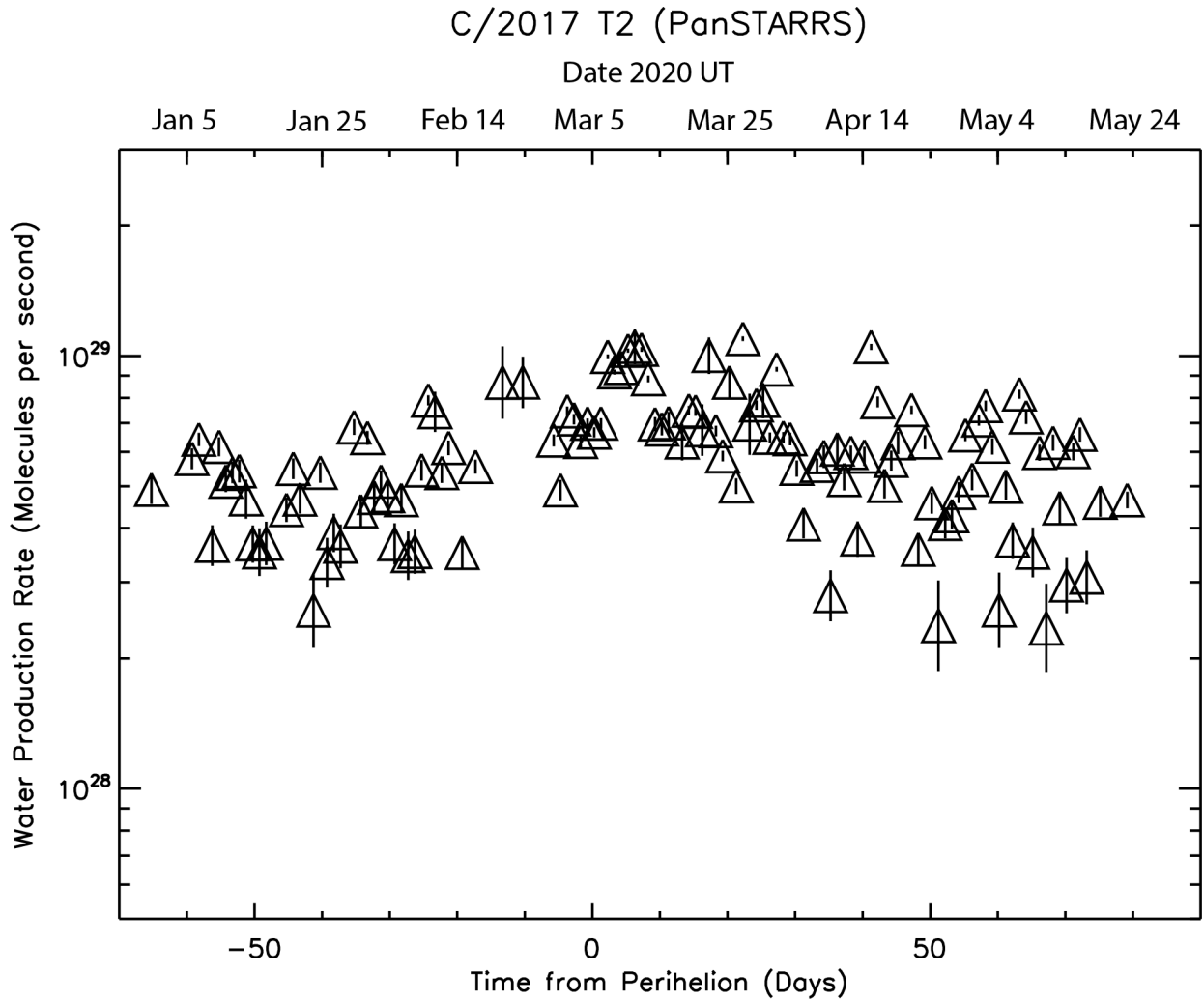

Figure 5. Water production rate in comet C/2017 T2 (PanSTARRS) as a function of time from perihelion and UT date 2020. The points give the water production rate in s$^{-1}$ from single images. The error bars give the 1-σ formal random fitting errors for each value. There is a ~30% uncertainty from the model parameters and calibration.

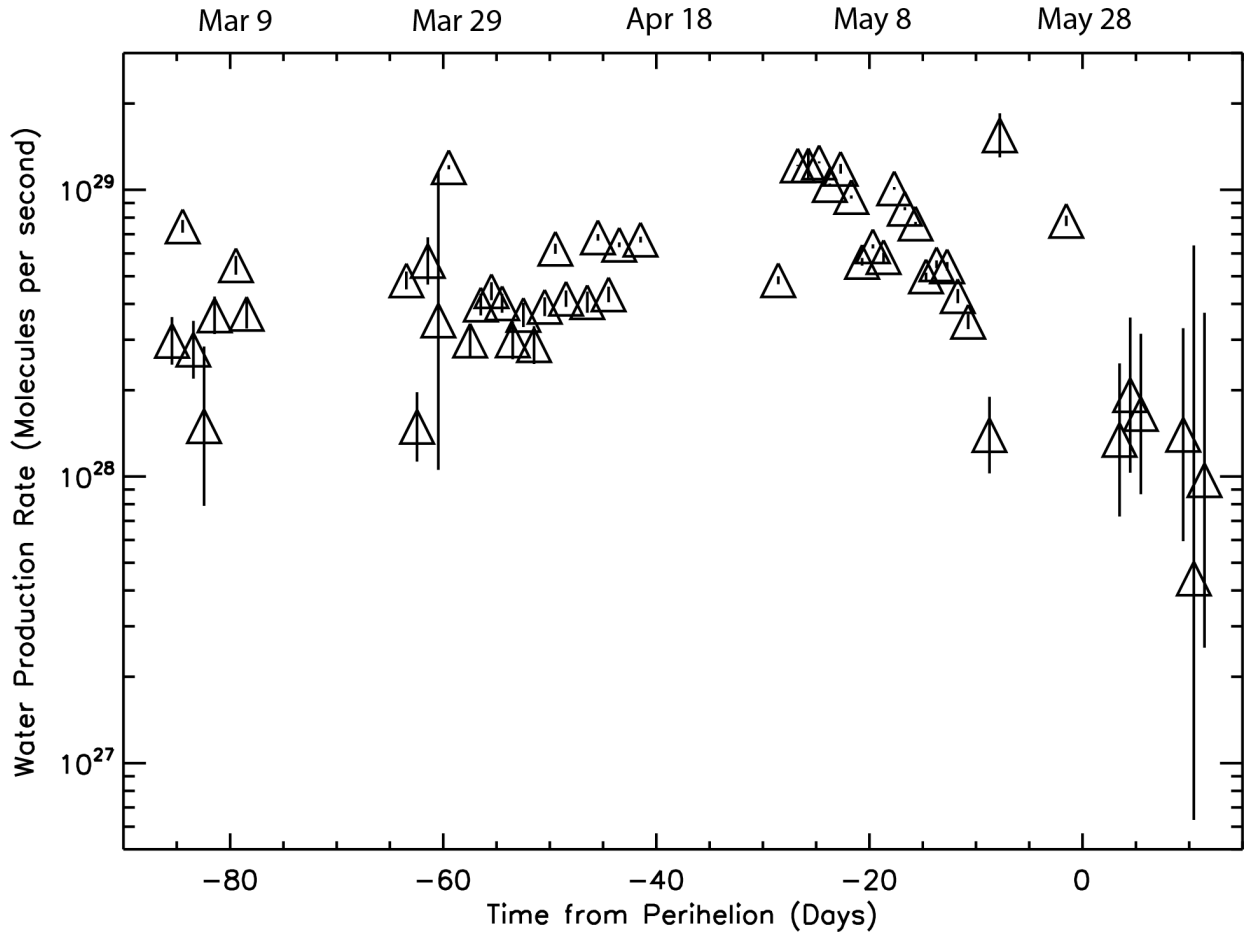

Figure 6. Water production rate in comet C/2020 F8 (SWAN) as a function of time from perihelion and UT date 2020. The points give the water production rate in s$^{-1}$ from single images. The error bars give the 1-σ formal random fitting errors for each value. There is a ~30% uncertainty from the model parameters and calibration.

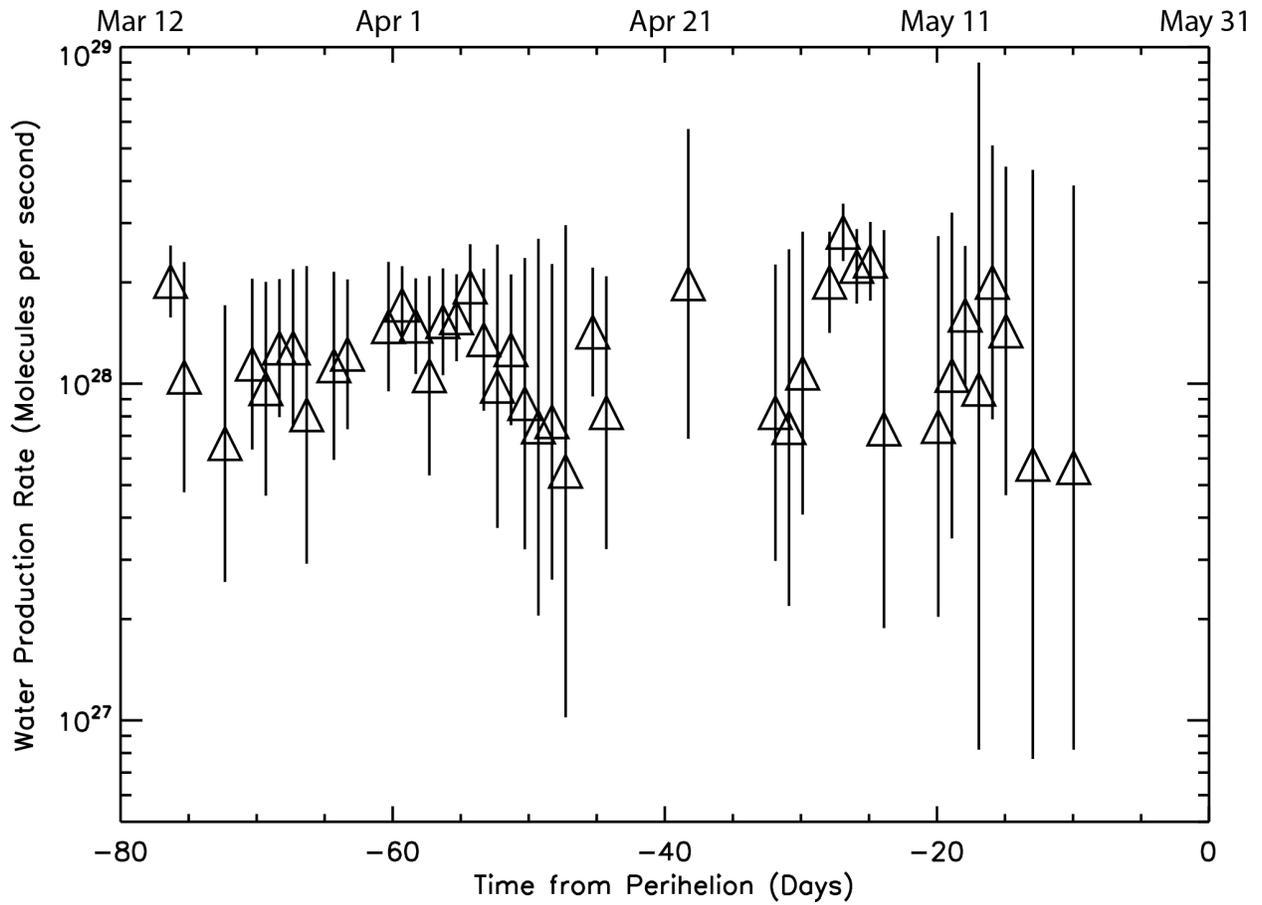

Figure 7. Water production rate in comet C/2019 Y4 (ATLAS) as a function of time from perihelion and UT date 2020. The points give the water production rate in s$^{-1}$ from single images. The error bars give the 1-σ formal random fitting errors for each value. There is a ~30% uncertainty from the model parameters and calibration.

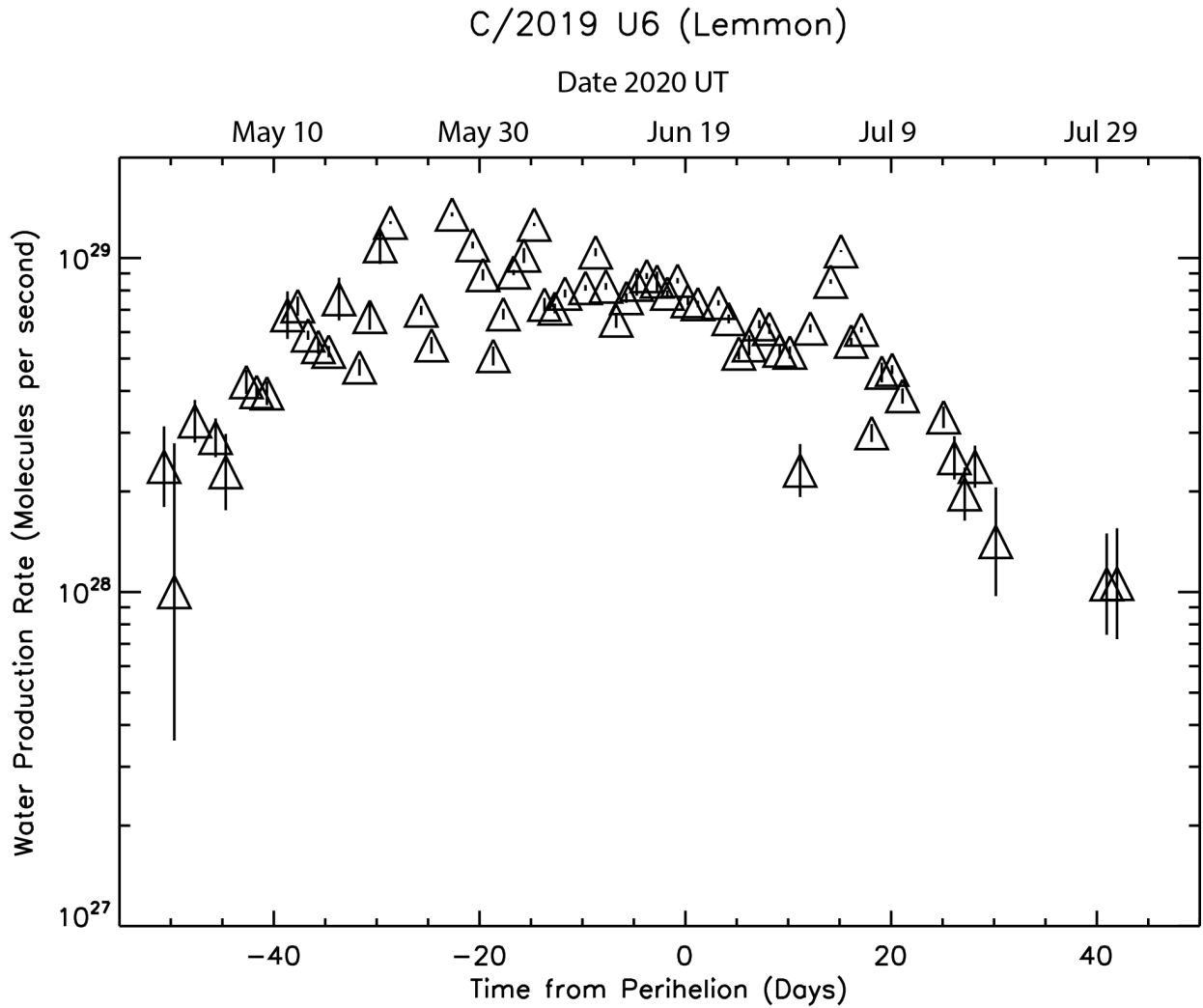

Figure 8. Water production rate in comet C/2019 U6 (Lemmon) as a function of time from perihelion and UT date 2020. The points give the water production rate in $s^{-1}$ from single images. The error bars give the 1-σ formal random fitting errors for each value. There is a ~30% uncertainty from the model parameters and calibration.

Table 1

Summary of SOHO/SWAN Observations

| Comet | Perihelion Date | q(au) | Images | $r_H$(au) |
|---|---|---|---|---|
| C/2015 V2 (Johnson) | 2017 Jun 12.34 | 1.637 | 64 | 1.638 – 1.758 |
| C/2019 Y1 (ATLAS) | 2020 Mar 15.56 | 0.837 | 57 | 0.838 – 1.627 |
| C/2017 T2 (PanSTARRS) | 2020 May 04.95 | 1.615 | 118 | 1.615 – 1.928 |
| C/2020 F8 (SWAN) | 2020 May 27.54 | 0.430 | 54 | 0.431 — 1.802 |
| C/2019 Y4 (ATLAS) | 2020 May 31.03 | 0.253 | 46 | 0.410 – 1.760 |
| C/2019 U6 (Lemmon) | 2020 Jun 18.82 | 0.914 | 68 | 0.914 – 1.270 |

Notes to Table 1

$r_H$ = heliocentric distance (au)

q = perihelion distance (au)

Table 2
SOHO/SWAN Observations of C/2015 V2 (Johnson) and Water Production Rates

| $\Delta T$ (Days) | r (au) | $\Delta$ (au) | g ($s^{-1}$) | Q ($10^{28}$ $s^{-1}$) | $\delta Q$ ($10^{28}$ $s^{-1}$) |
|---|---|---|---|---|---|
| −48.022 | 1.758 | 1.073 | 0.001474 | 1.64 | 1.14 |
| −47.022 | 1.754 | 1.063 | 0.001472 | 3.00 | 0.66 |
| −46.022 | 1.749 | 1.053 | 0.001475 | 4.57 | 0.46 |
| −44.692 | 1.743 | 1.040 | 0.001468 | 4.13 | 0.41 |
| −43.692 | 1.738 | 1.030 | 0.001471 | 3.56 | 0.56 |
| −42.692 | 1.734 | 1.020 | 0.001476 | 4.04 | 0.50 |
| −40.691 | 1.725 | 1.002 | 0.001504 | 3.39 | 0.53 |
| −39.691 | 1.721 | 0.993 | 0.001514 | 6.04 | 0.31 |
| −38.691 | 1.717 | 0.984 | 0.001524 | 3.34 | 0.53 |
| −37.691 | 1.713 | 0.975 | 0.001547 | 3.77 | 0.36 |
| −36.664 | 1.709 | 0.966 | 0.001540 | 3.08 | 0.57 |
| −35.664 | 1.705 | 0.957 | 0.001539 | 4.41 | 0.34 |
| −34.664 | 1.702 | 0.949 | 0.001521 | 4.00 | 0.41 |
| −33.664 | 1.698 | 0.941 | 0.001508 | 3.70 | 0.31 |
| −32.661 | 1.695 | 0.933 | 0.001495 | 4.03 | 0.82 |
| −31.661 | 1.691 | 0.925 | 0.001493 | 5.52 | 0.75 |
| −30.661 | 1.688 | 0.918 | 0.001468 | 3.99 | 0.75 |
| −29.635 | 1.685 | 0.911 | 0.001453 | 4.71 | 0.37 |
| −28.635 | 1.682 | 0.904 | 0.001444 | 4.20 | 0.41 |
| −26.632 | 1.676 | 0.890 | 0.001438 | 4.23 | 0.29 |
| −23.607 | 1.667 | 0.873 | 0.001446 | 3.49 | 0.42 |
| −20.603 | 1.660 | 0.858 | 0.001439 | 6.83 | 1.12 |
| −17.168 | 1.653 | 0.844 | 0.001455 | 5.93 | 0.50 |
| −16.174 | 1.651 | 0.841 | 0.001469 | 6.26 | 0.39 |
| −15.174 | 1.650 | 0.838 | 0.001494 | 5.48 | 0.59 |
| −14.196 | 1.648 | 0.835 | 0.001491 | 3.91 | 0.48 |
| −13.196 | 1.647 | 0.833 | 0.001493 | 4.97 | 0.47 |
| −12.203 | 1.645 | 0.832 | 0.001498 | 6.18 | 0.21 |
| −11.203 | 1.644 | 0.830 | 0.001502 | 5.24 | 0.27 |
| −10.203 | 1.643 | 0.829 | 0.001496 | 4.12 | 0.30 |
| −9.225 | 1.642 | 0.829 | 0.001479 | 4.89 | 0.13 |
| −8.224 | 1.641 | 0.828 | 0.001470 | 6.59 | 0.22 |
| −7.232 | 1.640 | 0.829 | 0.001442 | 6.01 | 0.63 |
| −6.232 | 1.639 | 0.829 | 0.001427 | 9.57 | 0.22 |

| DT | r | Δ | g | Q | δQ |
|---|---|---|---|---|---|
| −5.253 | 1.638 | 0.830 | 0.001421 | 8.62 | 0.23 |
| −4.253 | 1.638 | 0.832 | 0.001427 | 6.31 | 0.34 |
| −3.261 | 1.638 | 0.833 | 0.001433 | 6.74 | 0.35 |
| −2.261 | 1.637 | 0.836 | 0.001433 | 4.68 | 0.44 |
| −1.261 | 1.637 | 0.838 | 0.001445 | 4.91 | 0.31 |
| −0.281 | 1.637 | 0.841 | 0.001461 | 6.36 | 0.22 |
| 0.719 | 1.637 | 0.844 | 0.001464 | 7.12 | 0.20 |
| 1.710 | 1.637 | 0.848 | 0.001453 | 8.99 | 0.17 |
| 2.710 | 1.637 | 0.852 | 0.001442 | 7.00 | 0.23 |
| 3.690 | 1.638 | 0.857 | 0.001440 | 5.26 | 0.37 |
| 4.690 | 1.638 | 0.862 | 0.001443 | 4.40 | 0.41 |
| 5.681 | 1.639 | 0.867 | 0.001450 | 6.59 | 0.33 |
| 6.681 | 1.639 | 0.873 | 0.001452 | 6.47 | 0.30 |
| 7.681 | 1.640 | 0.879 | 0.001456 | 6.56 | 0.26 |
| 8.662 | 1.641 | 0.885 | 0.001456 | 4.83 | 0.61 |
| 9.662 | 1.642 | 0.892 | 0.001473 | 4.04 | 0.42 |
| 10.652 | 1.643 | 0.899 | 0.001488 | 7.34 | 0.20 |
| 11.652 | 1.644 | 0.906 | 0.001502 | 8.18 | 0.19 |
| 12.652 | 1.646 | 0.914 | 0.001500 | 8.39 | 0.21 |
| 13.637 | 1.647 | 0.922 | 0.001485 | 9.09 | 0.22 |
| 14.637 | 1.649 | 0.931 | 0.001481 | 6.24 | 0.28 |
| 15.595 | 1.650 | 0.939 | 0.001472 | 6.20 | 0.33 |
| 27.599 | 1.678 | 1.065 | 0.001416 | 5.50 | 0.42 |
| 28.599 | 1.681 | 1.077 | 0.001423 | 4.36 | 0.91 |
| 29.614 | 1.685 | 1.089 | 0.001431 | 3.88 | 0.74 |
| 30.614 | 1.688 | 1.102 | 0.001447 | 5.56 | 0.57 |
| 31.614 | 1.691 | 1.114 | 0.001455 | 3.93 | 0.66 |
| 32.614 | 1.694 | 1.127 | 0.001455 | 2.32 | 1.07 |
| 33.620 | 1.698 | 1.140 | 0.001471 | 1.79 | 1.34 |
| 34.620 | 1.702 | 1.153 | 0.001483 | 1.19 | 1.98 |

Notes. DT (Days from Perihelion June 12.34, 2017)

r: Heliocentric distance (au)

Δ: Comet-SOHO distance (au)

g: Solar Lyman-α g-factor (photons s$^{-1}$) at 1 au

Q: Water production rates for each image (s$^{-1}$)

δQ: internal 1-sigma uncertainties

Table 3

SOHO/SWAN Observations of C/2019 Y1 (ATLAS) and Water Production Rates

| $\Delta T$ (Days) | r (au) | $\Delta$ (au) | g ($s^{-1}$) | Q ($10^{28}$ $s^{-1}$) | $\delta Q$ ($10^{28}$ $s^{-1}$) |
|---|---|---|---|---|---|
| −14.704 | 0.882 | 1.576 | 0.001434 | 2.23 | 1.03 |
| −11.704 | 0.866 | 1.559 | 0.001397 | 1.25 | 1.68 |
| −10.709 | 0.861 | 1.552 | 0.001394 | 0.59 | 3.22 |
| −9.708 | 0.857 | 1.546 | 0.001388 | 0.83 | 2.71 |
| −8.708 | 0.853 | 1.539 | 0.001373 | 2.71 | 0.89 |
| −7.733 | 0.850 | 1.532 | 0.001378 | 3.59 | 0.73 |
| −6.733 | 0.847 | 1.525 | 0.001379 | 1.71 | 1.63 |
| −5.733 | 0.845 | 1.517 | 0.001382 | 2.13 | 1.10 |
| −3.737 | 0.841 | 1.502 | 0.001383 | 11.06 | 0.33 |
| −2.737 | 0.839 | 1.494 | 0.001381 | 20.38 | 0.15 |
| −1.737 | 0.838 | 1.486 | 0.001498 | 17.44 | 0.15 |
| 1.238 | 0.838 | 1.460 | 0.001373 | 1.98 | 1.21 |
| 8.206 | 0.852 | 1.395 | 0.001363 | 3.41 | 0.54 |
| 9.206 | 0.855 | 1.385 | 0.001364 | 3.24 | 0.59 |
| 10.206 | 0.859 | 1.375 | 0.001361 | 1.62 | 0.79 |
| 11.181 | 0.863 | 1.365 | 0.001368 | 2.51 | 0.91 |
| 12.181 | 0.868 | 1.355 | 0.001372 | 2.67 | 0.73 |
| 13.180 | 0.873 | 1.345 | 0.001369 | 3.86 | 0.60 |
| 18.152 | 0.903 | 1.295 | 0.001382 | 1.40 | 0.81 |
| 19.152 | 0.911 | 1.285 | 0.001394 | 1.75 | 1.13 |
| 20.148 | 0.918 | 1.275 | 0.001397 | 2.39 | 0.75 |
| 21.149 | 0.925 | 1.265 | 0.001417 | 1.72 | 1.44 |
| 22.149 | 0.933 | 1.256 | 0.001428 | 6.40 | 0.31 |
| 24.123 | 0.950 | 1.237 | 0.001445 | 8.17 | 0.24 |
| 25.123 | 0.958 | 1.228 | 0.001401 | 2.66 | 0.52 |
| 26.120 | 0.967 | 1.219 | 0.001408 | 3.70 | 0.38 |
| 37.067 | 1.078 | 1.138 | 0.001462 | 1.82 | 0.53 |
| 40.067 | 1.112 | 1.124 | 0.001471 | 0.89 | 1.34 |
| 44.694 | 1.166 | 1.111 | 0.001472 | 2.09 | 0.33 |
| 47.701 | 1.202 | 1.109 | 0.001454 | 1.77 | 0.48 |
| 48.701 | 1.214 | 1.109 | 0.001462 | 1.52 | 0.36 |
| 49.693 | 1.227 | 1.110 | 0.001475 | 2.73 | 0.25 |
| 50.693 | 1.239 | 1.112 | 0.001491 | 2.89 | 0.26 |
| 51.693 | 1.251 | 1.114 | 0.001487 | 2.87 | 0.24 |
| 52.693 | 1.264 | 1.117 | 0.001506 | 2.33 | 0.30 |
| 53.692 | 1.276 | 1.121 | 0.001500 | 2.81 | 0.22 |
| 54.692 | 1.289 | 1.125 | 0.001501 | 1.85 | 0.30 |
| 55.692 | 1.302 | 1.129 | 0.001513 | 2.15 | 0.26 |
| 56.674 | 1.314 | 1.135 | 0.001497 | 1.46 | 0.60 |
| 57.674 | 1.327 | 1.141 | 0.001509 | 1.62 | 0.38 |
| 58.674 | 1.340 | 1.147 | 0.001513 | 1.49 | 0.49 |
| 59.674 | 1.353 | 1.155 | 0.001504 | 1.56 | 0.39 |
| 60.663 | 1.365 | 1.163 | 0.001519 | 1.38 | 0.51 |
| 61.663 | 1.378 | 1.171 | 0.001522 | 1.76 | 0.43 |
| 64.646 | 1.417 | 1.200 | 0.001507 | 1.53 | 0.53 |
| 65.646 | 1.430 | 1.211 | 0.001523 | 0.93 | 0.63 |
| 66.634 | 1.443 | 1.222 | 0.001504 | 1.49 | 0.47 |
| 67.634 | 1.456 | 1.234 | 0.001504 | 1.74 | 0.59 |
| 70.618 | 1.495 | 1.273 | 0.001497 | 3.68 | 0.29 |
| 71.618 | 1.509 | 1.287 | 0.001502 | 0.63 | 1.37 |
| 73.605 | 1.535 | 1.316 | 0.001497 | 0.11 | 8.92 |
| 75.605 | 1.561 | 1.348 | 0.001510 | 1.80 | 0.49 |
| 76.589 | 1.574 | 1.364 | 0.001514 | 1.52 | 0.59 |
| 77.589 | 1.587 | 1.381 | 0.001544 | 1.90 | 0.46 |
| 78.576 | 1.600 | 1.397 | 0.001531 | 2.36 | 0.40 |
| 79.576 | 1.614 | 1.415 | 0.001519 | 1.78 | 0.53 |
| 80.576 | 1.627 | 1.433 | 0.001519 | 1.67 | 0.58 |
| −14.704 | 0.882 | 1.576 | 0.001434 | 2.23 | 1.03 |
| −11.704 | 0.866 | 1.559 | 0.001397 | 1.25 | 1.68 |
| −10.709 | 0.861 | 1.552 | 0.001394 | 0.59 | 3.22 |
| −9.708 | 0.857 | 1.546 | 0.001388 | 0.83 | 2.71 |
| −8.708 | 0.853 | 1.539 | 0.001373 | 2.71 | 0.89 |
| −7.733 | 0.850 | 1.532 | 0.001378 | 3.59 | 0.73 |
| −6.733 | 0.847 | 1.525 | 0.001379 | 1.71 | 1.63 |
| −5.733 | 0.845 | 1.517 | 0.001382 | 2.13 | 1.10 |
| −3.737 | 0.841 | 1.502 | 0.001383 | 11.06 | 0.33 |
| −2.737 | 0.839 | 1.494 | 0.001381 | 20.38 | 0.15 |
| −1.737 | 0.838 | 1.486 | 0.001498 | 17.44 | 0.15 |
| 1.238 | 0.838 | 1.460 | 0.001373 | 1.98 | 1.21 |
| 8.206 | 0.852 | 1.395 | 0.001363 | 3.41 | 0.54 |
| 9.206 | 0.855 | 1.385 | 0.001364 | 3.24 | 0.59 |
| 10.206 | 0.859 | 1.375 | 0.001361 | 1.62 | 0.79 |
| 11.181 | 0.863 | 1.365 | 0.001368 | 2.51 | 0.91 |
| 12.181 | 0.868 | 1.355 | 0.001372 | 2.67 | 0.73 |
| 13.180 | 0.873 | 1.345 | 0.001369 | 3.86 | 0.60 |
| 18.152 | 0.903 | 1.295 | 0.001382 | 1.40 | 0.81 |
| 19.152 | 0.911 | 1.285 | 0.001394 | 1.75 | 1.13 |
| 20.148 | 0.918 | 1.275 | 0.001397 | 2.39 | 0.75 |
| 21.149 | 0.925 | 1.265 | 0.001417 | 1.72 | 1.44 |
| 22.149 | 0.933 | 1.256 | 0.001428 | 6.40 | 0.31 |
| 24.123 | 0.950 | 1.237 | 0.001445 | 8.17 | 0.24 |
| 25.123 | 0.958 | 1.228 | 0.001401 | 2.66 | 0.52 |
| 26.120 | 0.967 | 1.219 | 0.001408 | 3.70 | 0.38 |
| 37.067 | 1.078 | 1.138 | 0.001462 | 1.82 | 0.53 |
| 40.067 | 1.112 | 1.124 | 0.001471 | 0.89 | 1.34 |
| 44.694 | 1.166 | 1.111 | 0.001472 | 2.09 | 0.33 |
| 47.701 | 1.202 | 1.109 | 0.001454 | 1.77 | 0.48 |
| 48.701 | 1.214 | 1.109 | 0.001462 | 1.52 | 0.36 |
| 49.693 | 1.227 | 1.110 | 0.001475 | 2.73 | 0.25 |
| 50.693 | 1.239 | 1.112 | 0.001491 | 2.89 | 0.26 |
| 51.693 | 1.251 | 1.114 | 0.001487 | 2.87 | 0.24 |
| 52.693 | 1.264 | 1.117 | 0.001506 | 2.33 | 0.30 |
| 53.692 | 1.276 | 1.121 | 0.001500 | 2.81 | 0.22 |
| 54.692 | 1.289 | 1.125 | 0.001501 | 1.85 | 0.30 |
| 55.692 | 1.302 | 1.129 | 0.001513 | 2.15 | 0.26 |
| 56.674 | 1.314 | 1.135 | 0.001497 | 1.46 | 0.60 |
| 57.674 | 1.327 | 1.141 | 0.001509 | 1.62 | 0.38 |
| 58.674 | 1.340 | 1.147 | 0.001513 | 1.49 | 0.49 |
| 59.674 | 1.353 | 1.155 | 0.001504 | 1.56 | 0.39 |
| 60.663 | 1.365 | 1.163 | 0.001519 | 1.38 | 0.51 |
| 61.663 | 1.378 | 1.171 | 0.001522 | 1.76 | 0.43 |
| 64.646 | 1.417 | 1.200 | 0.001507 | 1.53 | 0.53 |
| 65.646 | 1.430 | 1.211 | 0.001523 | 0.93 | 0.63 |
| 66.634 | 1.443 | 1.222 | 0.001504 | 1.49 | 0.47 |
| 67.634 | 1.456 | 1.234 | 0.001504 | 1.74 | 0.59 |
| 70.618 | 1.495 | 1.273 | 0.001497 | 3.68 | 0.29 |
| 71.618 | 1.509 | 1.287 | 0.001502 | 0.63 | 1.37 |
| 73.605 | 1.535 | 1.316 | 0.001497 | 0.11 | 8.92 |
| 75.605 | 1.561 | 1.348 | 0.001510 | 1.80 | 0.49 |
| 76.589 | 1.574 | 1.364 | 0.001514 | 1.52 | 0.59 |
| 77.589 | 1.587 | 1.381 | 0.001544 | 1.90 | 0.46 |

| DT | r | Δ | g | Q | δQ |
|---|---|---|---|---|---|
| 78.576 | 1.600 | 1.397 | 0.001531 | 2.36 | 0.40 |
| 79.576 | 1.614 | 1.415 | 0.001519 | 1.78 | 0.53 |
| 80.576 | 1.627 | 1.433 | 0.001519 | 1.67 | 0.58 |

Notes. DT (Days from Perihelion March 15.56, 2020)

r: Heliocentric distance (au)

Δ: Comet–SOHO distance (au)

g: Solar Lyman-α g-factor (photons s$^{-1}$) at 1 au

Q: Water production rates for each image (s$^{-1}$)

δQ: internal 1-sigma uncertainties

Table 4
SOHO/SWAN Observations of C/2017 T2 PanSTARRS and Water Production Rates

| ΔT (Days) | r (au) | Δ (au) | g (s$^{-1}$) | Q (10$^{28}$ s$^{-1}$) | δQ (10$^{28}$ s$^{-1}$) |
|---|---|---|---|---|---|
| −65.243 | 1.836 | 1.747 | 0.001431 | 4.91 | 0.39 |
| −59.243 | 1.800 | 1.761 | 0.001399 | 5.79 | 0.33 |
| −58.243 | 1.794 | 1.762 | 0.001407 | 6.41 | 0.23 |
| −56.243 | 1.783 | 1.766 | 0.001417 | 3.65 | 0.42 |
| −55.242 | 1.777 | 1.767 | 0.001485 | 6.17 | 0.32 |
| −54.242 | 1.772 | 1.769 | 0.001422 | 5.13 | 0.30 |
| −53.242 | 1.766 | 1.77 | 0.001422 | 5.35 | 0.28 |
| −52.242 | 1.761 | 1.771 | 0.001541 | 5.42 | 0.33 |
| −51.242 | 1.756 | 1.772 | 0.001411 | 4.67 | 0.51 |
| −50.260 | 1.751 | 1.773 | 0.001415 | 3.68 | 0.38 |
| −49.260 | 1.746 | 1.774 | 0.001416 | 3.52 | 0.48 |
| −48.260 | 1.741 | 1.774 | 0.001408 | 3.69 | 0.45 |
| −45.260 | 1.726 | 1.776 | 0.001396 | 4.41 | 0.29 |
| −44.260 | 1.721 | 1.776 | 0.001404 | 5.49 | 0.31 |
| −43.260 | 1.717 | 1.776 | 0.001403 | 4.66 | 0.36 |
| −41.260 | 1.708 | 1.776 | 0.001402 | 2.58 | 0.57 |
| −40.260 | 1.704 | 1.775 | 0.001404 | 5.38 | 0.30 |
| −39.260 | 1.700 | 1.775 | 0.001398 | 3.33 | 0.47 |
| −38.260 | 1.695 | 1.774 | 0.001401 | 3.90 | 0.42 |
| −37.260 | 1.691 | 1.774 | 0.001409 | 3.63 | 0.45 |
| −35.270 | 1.684 | 1.772 | 0.001394 | 6.85 | 0.29 |
| −34.270 | 1.680 | 1.771 | 0.001401 | 4.38 | 0.37 |
| −33.270 | 1.676 | 1.77 | 0.001387 | 6.46 | 0.24 |
| −32.269 | 1.673 | 1.769 | 0.001394 | 4.70 | 0.35 |
| −31.270 | 1.669 | 1.768 | 0.001392 | 5.13 | 0.31 |
| −30.269 | 1.666 | 1.767 | 0.001415 | 4.75 | 0.33 |
| −29.270 | 1.663 | 1.765 | 0.001411 | 3.68 | 0.43 |
| −28.269 | 1.660 | 1.764 | 0.001394 | 4.65 | 0.35 |
| −27.269 | 1.657 | 1.762 | 0.001433 | 3.45 | 0.47 |
| −26.269 | 1.654 | 1.76 | 0.001374 | 3.53 | 0.44 |
| −25.269 | 1.651 | 1.759 | 0.001376 | 5.44 | 0.30 |
| −24.269 | 1.648 | 1.757 | 0.001373 | 7.90 | 0.21 |
| −23.269 | 1.645 | 1.755 | 0.001384 | 7.43 | 0.84 |
| −22.269 | 1.643 | 1.753 | 0.001379 | 5.37 | 0.29 |
| −21.269 | 1.640 | 1.751 | 0.001394 | 6.13 | 0.24 |
| −19.269 | 1.636 | 1.746 | 0.001366 | 3.51 | 0.32 |
| −17.289 | 1.632 | 1.742 | 0.001374 | 5.55 | 0.21 |
| −13.289 | 1.625 | 1.732 | 0.001382 | 8.68 | 1.84 |
| −10.289 | 1.621 | 1.725 | 0.001382 | 8.69 | 1.27 |
| −5.723 | 1.617 | 1.714 | 0.001370 | 6.38 | 0.20 |
| −4.722 | 1.616 | 1.711 | 0.001367 | 4.90 | 0.27 |
| −3.723 | 1.616 | 1.708 | 0.001368 | 7.45 | 0.19 |
| −2.722 | 1.616 | 1.706 | 0.001347 | 7.14 | 0.20 |
| −1.723 | 1.615 | 1.703 | 0.001353 | 6.33 | 0.22 |
| −0.722 | 1.615 | 1.701 | 0.001356 | 6.98 | 0.19 |
| 0.277 | 1.615 | 1.698 | 0.001363 | 6.66 | 0.18 |
| 1.278 | 1.615 | 1.696 | 0.001375 | 6.99 | 0.20 |
| 2.278 | 1.615 | 1.694 | 0.001371 | 9.96 | 0.13 |
| 3.278 | 1.616 | 1.691 | 0.001386 | 9.18 | 0.12 |
| 4.277 | 1.616 | 1.689 | 0.001380 | 9.38 | 0.13 |
| 5.278 | 1.617 | 1.687 | 0.001390 | 10.29 | 0.11 |
| 6.278 | 1.617 | 1.685 | 0.001373 | 10.51 | 1.03 |
| 7.278 | 1.618 | 1.683 | 0.001382 | 10.35 | 0.13 |
| 8.278 | 1.619 | 1.681 | 0.001382 | 8.85 | 0.15 |
| 9.277 | 1.620 | 1.679 | 0.001384 | 6.96 | 0.27 |
| 10.278 | 1.621 | 1.677 | 0.001376 | 6.77 | 0.23 |
| 11.278 | 1.622 | 1.676 | 0.001388 | 7.00 | 0.42 |
| 13.278 | 1.625 | 1.673 | 0.001378 | 6.32 | 0.64 |
| 14.277 | 1.627 | 1.671 | 0.001372 | 7.48 | 0.18 |
| 15.278 | 1.628 | 1.670 | 0.001377 | 7.44 | 0.17 |
| 16.277 | 1.630 | 1.669 | 0.001390 | 6.74 | 0.99 |
| 17.258 | 1.632 | 1.668 | 0.001375 | 10.01 | 1.02 |
| 18.277 | 1.634 | 1.668 | 0.001376 | 6.71 | 0.19 |
| 19.277 | 1.636 | 1.667 | 0.001378 | 5.87 | 0.17 |
| 20.277 | 1.638 | 1.667 | 0.001383 | 8.69 | 0.78 |
| 21.277 | 1.641 | 1.666 | 0.001367 | 5.00 | 0.21 |
| 22.277 | 1.643 | 1.666 | 0.001371 | 10.96 | 0.13 |
| 23.277 | 1.645 | 1.666 | 0.001361 | 6.95 | 1.23 |
| 24.277 | 1.648 | 1.667 | 0.001360 | 7.70 | 0.19 |
| 25.277 | 1.651 | 1.667 | 0.001377 | 7.88 | 0.77 |
| 26.258 | 1.654 | 1.668 | 0.001357 | 6.48 | 0.18 |
| 27.258 | 1.657 | 1.669 | 0.001374 | 9.31 | 0.12 |
| 28.258 | 1.660 | 1.670 | 0.001402 | 6.48 | 0.19 |
| 29.258 | 1.663 | 1.671 | 0.001391 | 6.43 | 0.22 |
| 30.258 | 1.666 | 1.673 | 0.001382 | 5.49 | 0.24 |
| 31.258 | 1.669 | 1.674 | 0.001382 | 4.08 | 0.32 |
| 33.258 | 1.676 | 1.678 | 0.001378 | 5.58 | 0.22 |
| 34.259 | 1.680 | 1.681 | 0.001375 | 5.85 | 0.36 |
| 35.249 | 1.684 | 1.683 | 0.001380 | 2.79 | 0.41 |
| 36.249 | 1.687 | 1.686 | 0.001392 | 6.08 | 0.48 |
| 37.248 | 1.691 | 1.689 | 0.001373 | 5.17 | 0.30 |
| 38.249 | 1.695 | 1.693 | 0.001385 | 5.98 | 0.22 |
| 39.248 | 1.699 | 1.696 | 0.001378 | 3.77 | 0.37 |
| 40.249 | 1.704 | 1.700 | 0.001394 | 5.86 | 0.30 |
| 41.249 | 1.708 | 1.704 | 0.001402 | 10.49 | 0.17 |
| 42.230 | 1.712 | 1.709 | 0.001410 | 7.83 | 0.23 |
| 43.230 | 1.717 | 1.713 | 0.001405 | 5.01 | 0.35 |
| 44.230 | 1.721 | 1.718 | 0.001405 | 5.73 | 0.31 |
| 45.231 | 1.726 | 1.723 | 0.001400 | 6.29 | 0.38 |
| 47.231 | 1.736 | 1.735 | 0.001389 | 7.51 | 0.17 |
| 48.231 | 1.740 | 1.741 | 0.001379 | 3.57 | 0.32 |
| 49.220 | 1.745 | 1.747 | 0.001358 | 6.32 | 0.24 |
| 50.220 | 1.750 | 1.753 | 0.001368 | 4.57 | 0.26 |
| 51.220 | 1.756 | 1.760 | 0.001363 | 2.38 | 0.65 |
| 52.220 | 1.761 | 1.767 | 0.001384 | 4.10 | 0.33 |
| 53.220 | 1.766 | 1.774 | 0.001380 | 4.27 | 0.30 |
| 54.202 | 1.771 | 1.782 | 0.001386 | 4.83 | 0.27 |
| 55.202 | 1.777 | 1.790 | 0.001390 | 6.56 | 0.21 |
| 56.202 | 1.782 | 1.798 | 0.001380 | 5.18 | 0.30 |
| 57.202 | 1.788 | 1.806 | 0.001389 | 7.10 | 0.21 |
| 58.191 | 1.794 | 1.815 | 0.001386 | 7.67 | 0.21 |
| 59.191 | 1.799 | 1.824 | 0.001391 | 6.18 | 0.27 |
| 60.191 | 1.805 | 1.833 | 0.001381 | 2.58 | 0.57 |
| 61.191 | 1.811 | 1.842 | 0.001383 | 4.99 | 0.36 |
| 62.191 | 1.817 | 1.852 | 0.001385 | 3.74 | 0.38 |
| 63.191 | 1.823 | 1.862 | 0.001378 | 8.16 | 0.20 |
| 64.191 | 1.829 | 1.872 | 0.001385 | 7.22 | 0.27 |
| 65.173 | 1.835 | 1.882 | 0.001396 | 3.51 | 0.50 |
| 66.173 | 1.842 | 1.893 | 0.001406 | 5.98 | 0.26 |
| 67.173 | 1.848 | 1.904 | 0.001412 | 2.35 | 0.63 |

| DT | r | Δ | g | Q | δQ |
|---|---|---|---|---|---|
| 68.173 | 1.854 | 1.915 | 0.001401 | 6.30 | 0.28 |
| 69.173 | 1.861 | 1.927 | 0.001425 | 4.45 | 0.39 |
| 70.173 | 1.867 | 1.938 | 0.001423 | 2.95 | 0.47 |
| 71.163 | 1.874 | 1.950 | 0.001405 | 6.01 | 0.28 |
| 72.162 | 1.880 | 1.962 | 0.001390 | 6.58 | 0.25 |
| 73.163 | 1.887 | 1.974 | 0.001419 | 3.08 | 0.47 |
| 75.163 | 1.901 | 2.000 | 0.001402 | 4.58 | 0.36 |
| 79.145 | 1.928 | 2.053 | 0.001412 | 4.65 | 0.21 |

Notes. DT (Days from Perihelion May 4.95, 2020)

r: Heliocentric distance (au)

Δ: Comet-SOHO distance (au)

g: Solar Lyman-α g-factor (photons $s^{-1}$) at 1 au

Q: Water production rates for each image ($s^{-1}$)

δQ: internal 1-sigma uncertainties

Table 5
SOHO/SWAN Observations of C/2020 F8 (SWAN) and Water Production Rates

| ΔT (Days) | r (au) | Δ (au) | g (s$^{-1}$) | Q (10$^{28}$ s$^{-1}$) | δQ (10$^{28}$ s$^{-1}$) |
|---|---|---|---|---|---|
| −85.446 | 1.802 | 2.336 | 0.001659 | 2.97 | 0.62 |
| −84.446 | 1.786 | 2.314 | 0.001698 | 7.46 | 0.39 |
| −83.446 | 1.770 | 2.291 | 0.001672 | 2.77 | 0.72 |
| −82.446 | 1.755 | 2.269 | 0.001733 | 1.50 | 1.34 |
| −81.446 | 1.739 | 2.246 | 0.001666 | 3.65 | 0.59 |
| −79.446 | 1.706 | 2.199 | 0.001676 | 5.45 | 0.42 |
| −78.446 | 1.690 | 2.175 | 0.001680 | 3.69 | 0.46 |
| −63.452 | 1.442 | 1.796 | 0.001708 | 4.83 | 0.36 |
| −62.452 | 1.425 | 1.769 | 0.002645 | 1.49 | 0.48 |
| −61.452 | 1.408 | 1.742 | 0.001728 | 5.65 | 1.18 |
| −60.452 | 1.391 | 1.714 | 0.001810 | 3.50 | 8.10 |
| −59.474 | 1.374 | 1.687 | 0.001745 | 11.99 | 0.19 |
| −57.473 | 1.339 | 1.632 | 0.001888 | 2.98 | 0.41 |
| −56.474 | 1.322 | 1.604 | 0.001734 | 3.98 | 0.36 |
| −55.473 | 1.305 | 1.575 | 0.001737 | 4.43 | 0.33 |
| −54.473 | 1.287 | 1.547 | 0.001749 | 4.02 | 0.32 |
| −53.473 | 1.270 | 1.518 | 0.001738 | 2.98 | 0.48 |
| −52.473 | 1.252 | 1.489 | 0.001742 | 3.66 | 0.38 |
| −51.473 | 1.234 | 1.460 | 0.001733 | 2.87 | 0.47 |
| −50.473 | 1.217 | 1.431 | 0.001748 | 3.91 | 0.30 |
| −49.473 | 1.199 | 1.402 | 0.001753 | 6.23 | 0.25 |
| −48.473 | 1.181 | 1.372 | 0.001748 | 4.17 | 0.28 |
| −46.481 | 1.145 | 1.313 | 0.001746 | 4.06 | 0.36 |
| −45.481 | 1.127 | 1.283 | 0.001762 | 6.83 | 0.18 |
| −44.481 | 1.109 | 1.253 | 0.001765 | 4.32 | 0.28 |
| −43.481 | 1.091 | 1.223 | 0.001764 | 6.43 | 0.12 |
| −41.481 | 1.055 | 1.163 | 0.001792 | 6.70 | 0.16 |
| −28.564 | 0.817 | 0.776 | 0.001764 | 4.84 | 0.15 |
| −26.737 | 0.783 | 0.726 | 0.001759 | 12.16 | 0.06 |
| −25.736 | 0.764 | 0.700 | 0.001779 | 12.18 | 1.47 |
| −24.728 | 0.746 | 0.675 | 0.001774 | 12.47 | 0.09 |
| −23.728 | 0.727 | 0.652 | 0.001794 | 10.39 | 0.13 |
| −22.709 | 0.709 | 0.629 | 0.001772 | 11.84 | 0.46 |
| −21.703 | 0.691 | 0.608 | 0.001751 | 9.44 | 0.12 |

| ΔT | r | Δ | g | Q | δQ |
|---|---|---|---|---|---|
| −20.684 | 0.672 | 0.590 | 0.001772 | 5.61 | 0.17 |
| −19.680 | 0.655 | 0.574 | 0.001747 | 6.35 | 0.10 |
| −18.680 | 0.637 | 0.560 | 0.001757 | 5.82 | 0.21 |
| −17.680 | 0.620 | 0.550 | 0.001749 | 10.12 | 0.10 |
| −16.680 | 0.603 | 0.543 | 0.001740 | 8.60 | 0.11 |
| −15.677 | 0.586 | 0.540 | 0.001711 | 7.63 | 0.10 |
| −14.691 | 0.570 | 0.540 | 0.001695 | 4.99 | 0.15 |
| −13.710 | 0.555 | 0.543 | 0.001695 | 5.50 | 0.17 |
| −12.710 | 0.540 | 0.551 | 0.001673 | 5.41 | 0.19 |
| −11.713 | 0.525 | 0.563 | 0.001637 | 4.26 | 0.25 |
| −10.740 | 0.511 | 0.578 | 0.001605 | 3.47 | 0.21 |
| −8.742 | 0.486 | 0.618 | 0.001557 | 1.39 | 0.50 |
| −7.769 | 0.475 | 0.642 | 0.001526 | 15.50 | 2.99 |
| −1.530 | 0.431 | 0.847 | 0.001387 | 7.80 | 0.33 |
| 3.468 | 0.439 | 1.039 | 0.001387 | 1.34 | 1.14 |
| 4.468 | 0.445 | 1.077 | 0.001379 | 1.92 | 1.66 |
| 5.468 | 0.453 | 1.115 | 0.001414 | 1.65 | 1.49 |
| 9.441 | 0.494 | 1.260 | 0.001503 | 1.40 | 1.89 |
| 10.441 | 0.507 | 1.294 | 0.001513 | 0.44 | 5.95 |
| 11.440 | 0.521 | 1.328 | 0.001530 | 0.97 | 2.75 |

Notes. ΔT (Days from Perihelion May 27.54, 2020)

r: Heliocentric distance (au)

Δ: SOHO distance (au)

g: Solar Lyman-α g-factor (photons $s^{-1}$) at 1 au

Q: Water production rates for each image ($s^{-1}$)

δQ: internal 1-sigma uncertainties

Table 6
SOHO/SWAN Observations of C/2019 Y4 (ATLAS) and Water Production Rates

| $\Delta T$ (Days) | $r$ (au) | $\Delta$ (au) | $g$ ($s^{-1}$) | $Q$ ($10^{28}$ $s^{-1}$) | $\delta Q$ ($10^{28}$ $s^{-1}$) |
|---|---|---|---|---|---|
| −76.320 | 1.760 | 1.124 | 0.002647 | 2.01 | 0.56 |
| −75.320 | 1.743 | 1.119 | 0.002647 | 1.05 | 1.25 |
| −72.320 | 1.692 | 1.103 | 0.002646 | 0.66 | 1.05 |
| −70.320 | 1.657 | 1.094 | 0.002680 | 1.14 | 0.91 |
| −69.321 | 1.640 | 1.090 | 0.002680 | 0.97 | 1.04 |
| −68.321 | 1.623 | 1.085 | 0.002680 | 1.28 | 0.77 |
| −67.321 | 1.605 | 1.081 | 0.002679 | 1.28 | 0.91 |
| −66.321 | 1.587 | 1.077 | 0.002679 | 0.81 | 1.43 |
| −64.321 | 1.552 | 1.070 | 0.002679 | 1.13 | 1.02 |
| −63.321 | 1.534 | 1.066 | 0.002713 | 1.22 | 0.82 |
| −60.308 | 1.480 | 1.055 | 0.002712 | 1.48 | 0.82 |
| −59.308 | 1.461 | 1.052 | 0.002712 | 1.71 | 0.52 |
| −58.308 | 1.443 | 1.048 | 0.002712 | 1.48 | 0.57 |
| −57.308 | 1.425 | 1.045 | 0.002711 | 1.06 | 1.03 |
| −56.308 | 1.406 | 1.041 | 0.002745 | 1.53 | 0.67 |
| −55.308 | 1.388 | 1.038 | 0.002744 | 1.57 | 0.54 |
| −54.307 | 1.369 | 1.035 | 0.002744 | 1.94 | 0.66 |
| −53.307 | 1.350 | 1.031 | 0.002744 | 1.35 | 0.84 |
| −52.307 | 1.331 | 1.028 | 0.002744 | 0.98 | 1.61 |
| −51.307 | 1.312 | 1.024 | 0.002777 | 1.26 | 0.85 |
| −50.294 | 1.293 | 1.021 | 0.002777 | 0.87 | 1.49 |
| −49.294 | 1.274 | 1.017 | 0.002776 | 0.74 | 1.95 |
| −48.295 | 1.254 | 1.014 | 0.002776 | 0.77 | 1.50 |
| −47.295 | 1.235 | 1.010 | 0.002776 | 0.55 | 2.41 |
| −45.295 | 1.196 | 1.002 | 0.002808 | 1.42 | 0.79 |
| −44.295 | 1.176 | 0.998 | 0.002807 | 0.82 | 1.27 |
| −38.277 | 1.054 | 0.971 | 0.002837 | 1.98 | 3.73 |
| −31.873 | 0.919 | 0.937 | 0.002896 | 0.82 | 1.44 |
| −30.872 | 0.897 | 0.931 | 0.002896 | 0.74 | 1.77 |
| −29.872 | 0.875 | 0.925 | 0.002895 | 1.08 | 1.75 |
| −27.894 | 0.832 | 0.911 | 0.002924 | 2.00 | 0.83 |
| −26.894 | 0.809 | 0.904 | 0.002923 | 2.82 | 0.61 |
| −25.894 | 0.787 | 0.897 | 0.002952 | 2.23 | 0.65 |
| −24.894 | 0.764 | 0.890 | 0.002952 | 2.31 | 0.71 |

| DT | r | Δ | g | Q | δQ |
|---|---|---|---|---|---|
| −23.900 | 0.741 | 0.882 | 0.002951 | 0.73 | 2.13 |
| −22.900 | 0.718 | 0.875 | 0.002951 | 0.30 | 5.00 |
| −19.899 | 0.648 | 0.851 | 0.002977 | 0.75 | 2.00 |
| −18.899 | 0.625 | 0.843 | 0.003004 | 1.06 | 2.16 |
| −17.924 | 0.601 | 0.835 | 0.003004 | 1.60 | 0.97 |
| −15.928 | 0.554 | 0.819 | 0.003003 | 2.00 | 3.10 |
| −14.928 | 0.529 | 0.811 | 0.003002 | 1.44 | 2.98 |
| −12.953 | 0.482 | 0.796 | 0.003002 | 0.58 | 3.74 |
| −9.955 | 0.410 | 0.778 | 0.002973 | 0.56 | 3.31 |

Notes. DT (Days from Perihelion May 31.03, 2020)

r: Heliocentric distance (au)

Δ: Comet-SOHO distance (au)

g: Solar Lyman-α g-factor (photons $s^{-1}$) at 1 au

Q: Water production rates for each image ($s^{-1}$)

δQ: internal 1-sigma uncertainties

Table 7

SOHO/SWAN Observations of C/2019 U6 (Lemmon) and Water Production Rates

| $\Delta T$ (Days) | r (au) | $\Delta$ (au) | g ($s^{-1}$) | Q ($10^{28}$ $s^{-1}$) | $\delta Q$ ($10^{28}$ $s^{-1}$) |
|---|---|---|---|---|---|
| −50.663 | 1.270 | 1.566 | 0.001558 | 2.37 | 0.76 |
| −49.663 | 1.258 | 1.553 | 0.001536 | 1.00 | 1.79 |
| −47.663 | 1.236 | 1.526 | 0.001540 | 3.25 | 0.51 |
| −45.663 | 1.214 | 1.498 | 0.001542 | 2.90 | 0.41 |
| −44.663 | 1.203 | 1.484 | 0.001544 | 2.29 | 0.69 |
| −42.663 | 1.181 | 1.455 | 0.001526 | 4.26 | 0.36 |
| −41.663 | 1.171 | 1.441 | 0.001544 | 3.98 | 0.17 |
| −40.663 | 1.160 | 1.426 | 0.001533 | 3.94 | 0.33 |
| −38.663 | 1.140 | 1.397 | 0.001509 | 6.75 | 1.20 |
| −37.663 | 1.130 | 1.382 | 0.001503 | 7.18 | 0.48 |
| −36.663 | 1.120 | 1.367 | 0.001516 | 5.87 | 0.19 |
| −35.663 | 1.110 | 1.352 | 0.001481 | 5.42 | 0.27 |
| −34.663 | 1.100 | 1.336 | 0.001474 | 5.25 | 0.21 |
| −33.663 | 1.091 | 1.321 | 0.001482 | 7.54 | 1.20 |
| −31.663 | 1.073 | 1.291 | 0.001458 | 4.71 | 0.28 |
| −30.663 | 1.064 | 1.275 | 0.001461 | 6.68 | 0.63 |
| −29.662 | 1.055 | 1.260 | 0.001456 | 10.86 | 1.42 |
| −28.662 | 1.046 | 1.245 | 0.001471 | 12.77 | 0.13 |
| −25.668 | 1.022 | 1.198 | 0.001471 | 6.97 | 0.22 |
| −24.668 | 1.014 | 1.183 | 0.001474 | 5.48 | 0.32 |
| −22.668 | 1.000 | 1.153 | 0.001478 | 13.51 | 0.18 |
| −20.668 | 0.986 | 1.123 | 0.001450 | 10.94 | 0.25 |
| −19.668 | 0.979 | 1.108 | 0.001442 | 8.90 | 0.35 |
| −18.668 | 0.973 | 1.093 | 0.001438 | 5.09 | 0.34 |
| −17.691 | 0.967 | 1.079 | 0.001434 | 6.79 | 0.26 |
| −16.691 | 0.962 | 1.065 | 0.001434 | 9.06 | 0.17 |
| −15.691 | 0.956 | 1.050 | 0.001446 | 10.18 | 0.54 |
| −14.691 | 0.951 | 1.036 | 0.001423 | 12.60 | 0.13 |
| −13.690 | 0.947 | 1.023 | 0.001415 | 7.26 | 0.34 |
| −12.697 | 0.942 | 1.010 | 0.001427 | 7.06 | 0.24 |
| −11.697 | 0.938 | 0.996 | 0.001415 | 7.83 | 0.22 |
| −9.697 | 0.931 | 0.971 | 0.001432 | 8.14 | 0.16 |
| −8.719 | 0.928 | 0.960 | 0.001437 | 10.41 | 0.29 |
| −7.718 | 0.925 | 0.948 | 0.001426 | 8.22 | 0.19 |

| DT | r | Δ | | | |
|---|---|---|---|---|---|
| −6.718 | 0.922 | 0.937 | 0.001419 | 6.50 | 0.33 |
| −5.726 | 0.920 | 0.926 | 0.001415 | 7.59 | 0.25 |
| −4.726 | 0.918 | 0.916 | 0.001414 | 8.32 | 0.66 |
| −3.747 | 0.917 | 0.907 | 0.001389 | 8.83 | 0.18 |
| −2.747 | 0.916 | 0.897 | 0.001355 | 8.51 | 0.59 |
| −1.747 | 0.915 | 0.889 | 0.001382 | 7.83 | 0.17 |
| −0.755 | 0.914 | 0.881 | 0.001345 | 8.56 | 0.16 |
| 0.245 | 0.914 | 0.874 | 0.001362 | 7.44 | 0.21 |
| 1.224 | 0.915 | 0.867 | 0.001353 | 7.28 | 0.17 |
| 3.216 | 0.916 | 0.856 | 0.001374 | 7.34 | 0.14 |
| 4.216 | 0.917 | 0.851 | 0.001380 | 6.57 | 0.20 |
| 5.196 | 0.919 | 0.847 | 0.001380 | 5.18 | 0.22 |
| 6.196 | 0.921 | 0.844 | 0.001380 | 5.48 | 0.39 |
| 7.187 | 0.923 | 0.841 | 0.001366 | 6.34 | 0.18 |
| 8.167 | 0.926 | 0.839 | 0.001367 | 6.14 | 0.22 |
| 9.168 | 0.929 | 0.838 | 0.001378 | 5.28 | 0.25 |
| 10.159 | 0.932 | 0.838 | 0.001366 | 5.22 | 0.23 |
| 11.158 | 0.936 | 0.838 | 0.001375 | 2.31 | 0.46 |
| 12.139 | 0.940 | 0.840 | 0.001380 | 6.16 | 0.17 |
| 14.132 | 0.949 | 0.844 | 0.001372 | 8.50 | 0.12 |
| 15.132 | 0.954 | 0.847 | 0.001377 | 10.49 | 0.07 |
| 16.119 | 0.959 | 0.851 | 0.001371 | 5.63 | 0.14 |
| 17.112 | 0.964 | 0.856 | 0.001378 | 6.10 | 0.12 |
| 18.112 | 0.970 | 0.862 | 0.001395 | 2.99 | 0.19 |
| 19.099 | 0.976 | 0.868 | 0.001406 | 4.54 | 0.32 |
| 20.099 | 0.982 | 0.874 | 0.001411 | 4.63 | 0.14 |
| 21.099 | 0.989 | 0.881 | 0.001421 | 3.86 | 0.21 |
| 25.099 | 1.018 | 0.916 | 0.001415 | 3.34 | 0.25 |
| 26.150 | 1.026 | 0.927 | 0.001402 | 2.52 | 0.40 |
| 27.150 | 1.034 | 0.937 | 0.001436 | 1.96 | 0.39 |
| 28.157 | 1.042 | 0.948 | 0.001416 | 2.37 | 0.37 |
| 30.176 | 1.059 | 0.972 | 0.001400 | 1.41 | 0.64 |
| 40.957 | 1.163 | 1.119 | 0.001448 | 1.60 | 0.67 |
| 41.957 | 1.174 | 1.134 | 0.001447 | 1.61 | 0.75 |

Notes. DT (Days from Perihelion June 18.82, 2020)

r: Heliocentric distance (au)

Δ: Comet–SOHO distance (au)

g: Solar Lyman-$\alpha$ g-factor (photons s$^{-1}$) at 1 au

Q: Water production rates for each image (s$^{-1}$)

$\delta$Q: internal 1-sigma uncertainties